\documentclass[aps,prd,preprint,superscriptaddress,amsmath,amssymb,showpacs]{revtex4-1}
\usepackage{dcolumn}
\usepackage{graphicx}
\usepackage{float}
\usepackage{physics}
\usepackage[colorlinks=true,allcolors=blue]{hyperref}

\begin{document}
	
\title{Holographic deconfined QGP phase diagram and entropy with an anomalous flow in a magnetic field background}
	
\author{Jiali Deng}
\affiliation{College of Science, China Three Gorges University, Yichang 443002, China}
	
\author{Sheng-Qin Feng}
\email{fengsq@ctgu.edu.cn}
\affiliation{College of Science, China Three Gorges University, Yichang 443002, China}
\affiliation{Key Laboratory of Quark and Lepton Physics (MOE) and Institute of Particle Physics,\\
Central China Normal University, Wuhan 430079, China}

	
\date{\today}
	
\begin{abstract}
We assume that the initial hydrodynamic environment is a quark gluon plasma(QGP) phase where merely $u$ and $d$ quarks are considered when adding a magnetic field. When considering the chiral magnetic effect in relativistic heavy ion collisions, an anomalous current will be formed in the \textrm{QGP} environment. The chiral magnetic current formed by these $u$ and $d$ quarks has an impact on the heavy quarkonium. By using fluid/gravity duality, the metric with anomalous flow is established by using fluid/gravity duality, so as to introduce the magnetic field effect into the corresponding metric. And then we use heavy quarkonium as a probe to study phase transition, and utilize the effective string tension of the heavy quarkonium to study phase transition by AdS/QCD theory. The characteristics of reporting inverse magnetic catalysis for the confinement-deconfinement transition with anomalous flow are in qualitative agreement with lattice \textrm{QCD} findings. The heavy-quarkonium asymptotic entropy distributions with different magnetic field  around the confinement-deconfinement transition temperature are given in the paper.
\end{abstract}
	
\maketitle	

\section{Introduction}\label{sec:01_intro}
The dissociation of charmonium and bottomonium bound states has been proposed as a signal for the formation of a hot and deconfined quark-gluon plasma by Matsui and Satz~~\cite{Matsui:1986dk}. Quarkonium bound states such as charmonium $c\bar{c}$ and bottomonium $b\bar{b}$ will dissociate because of color screening in a deconfined \textrm{QGP} and thereby exhibit a suppression relative to the confined phase. Because of the importance of the problem, $c\bar{c}$ and $b\bar{b}$ dissociation in \textrm{QGP} have been a focus of many recent studies~~\cite{Rapp:2008tf}.

Some analyses of the $q\bar{q}$ dissociation of phenomenon are based on the study of the quark-antiquark static potential taken from lattice \textrm{QCD}~~\cite{Satz:2005hx}. Nevertheless, the created quarkonium bound state is not static, but has a velocity $\upsilon =\tanh(\eta)$ with respect to the medium in relativistic heavy ion collisions. If the relative speed of quarkonium exceeds a typical thermal speed, it can be considered that the quarkonium suppression is stronger than the thermal dissociation in a static heat bath \cite{Liu:2006nn,Feng:2019boe}. In the presence of an external magnetic field, the quarkonium dissociation in a parity-violating chiral plasma has been investigated \cite{Sadofyev:2015hxa}. This environment of \textrm{QGP} is an approximately chiral one which can be created in relativistic heavy-ion collisions.  When two relativistic heavy ions collide with a nonzero impact parameter, a strong magnetic field with a magnitude of the order of~~ \cite{Skokov:2009qp,Voronyuk:2011jd,Bzdak:2011yy,Deng:2012pc,Mo:2013qya,Zhong:2014cda,Feng:2016srp,Kharzeev:2007jp}  $eB\sim m_{\pi}^{2}$ ($m_{\pi}$ is the pion mass), is produced in the direction of the angular momentum of the collision.

The chirality imbalance should have experimental consequences in such a strong magnetic field. If the chirality is nonzero, the quark spins are locked either parallel or antiparallel to the magnetic field direction, depending on the quark charge. This would lead to a charge separation in the final state, and an electromagnetic current is generated along the direction of the magnetic field \cite{Deng:2012pc,Mo:2013qya,Zhong:2014cda,Feng:2016srp,Kharzeev:2007jp}. Anomaly-induced effects in a \textrm{QGP} medium, which is called chiral magnetic effect (CME) \cite{Kharzeev:2012ph,Zakharov:2012vv,Kharzeev:2013ffa,Liao:2014ava}, have attracted much attention \cite{Guo:2019joy,She:2017icp}.

It is well known that the deconfinement phase of \textrm{QCD} is at a higher temperature and density, while the confinement phase is at a lower temperature and density. The probe of the phase structure of \textrm{QCD} is an important and challenging assignment. It is generally realized that there is a phase transition between the two phases. How to study phase diagrams  in the $T-\mu$ plane is a rather hard job because the \textrm{QCD} coupling constant becomes very large near the phase change region, and the traditional perturbation \textrm{QCD} method can not be used. For a long time, the lattice \textrm{QCD} method is regarded as the only credible way to study the problem. Although lattice \textrm{QCD} works well for zero baryon density, but it has the sign problem when taking finite baryon density into account. However, the most interesting region in the \textrm{QCD} phase diagram is with a finite baryon density. The issue of finite baryon number density is the most concerned one in both relativistic heavy ion collisions and compact stars in astrophysics.

This situation was greatly improved with the advent of the AdS/CFT correspondence to stimulate the interest again in finding a string description of the strong interactions. Its developing theory called AdS/QCD used a five-dimensional effective description to study \textrm{QCD}. The effective string tension was used to discuss the thermal phase transition characteristics of a static $q\bar{q}$ pair \cite{Andreev:2006nw}. And then it was extended to discuss the thermal phase transition of the moving quarkonium \cite{Chen:2017lsf}. Reference \cite{Chen:2017lsf,Bohra:2019ebj} utilized effective string tension to study confinement-deconfinement phase transition with a background magnetic field.

The effects of a background magnetic field on the chiral condensate and confinement-deconfinement transition have been investigated by some publications~~\cite{Rodrigues:2017cha,Rodrigues:2017iqi,Rodrigues:2018pep,McInnes:2015kec,Gursoy:2017wzz,Gursoy:2016ofp,Dudal:2018rki}. Sensible $2+1$ dimensions gravity solutions manifesting inverse magnetic catalysis have been investigated in \cite{Rodrigues:2017cha,Rodrigues:2017iqi,Rodrigues:2018pep}, and $3+1$ dimensions confinement-deconfinement phase transition revealing the possibility of inverse catalysis have been studied in Refs. \cite{Gursoy:2017wzz,Gursoy:2016ofp}.

The fluid/gravity correspondence \cite{Bhattacharyya:2007vjd} is a very powerful tool to investigate the hydrodynamic regime of quantum field theories with holographic dual. This skill has been dedicated to the study of the positivity of the entropy production using methods of black hole thermodynamics \cite{Bhattacharyya:2008xc,Loganayagam:2008is,Chapman:2012my}. It is also very useful for the calculation of transport coefficients. The application of the fluid/gravity correspondence to theories including chiral anomalies \cite{Erdmenger:2008rm,Banerjee:2008th} lead to study chiral magnetic effects.  In this paper, we suppose that the initial hydrodynamic environment is a \textrm{QGP} phase where merely $u$ and $d$ quarks are considered when adding a magnetic field. The quark gluon plasma produced by relativistic heavy ion collision will produce anomalous flow. The chiral magnetic current formed by these $u$ and $d$ quarks has an impact on the heavy quarkonium. And then we use heavy quarkonium as a probe to study phase transition. Specifically, the effective string tension method is used to discuss the phase transition by AdS/QCD duality theory.

The paper is organized as follows. In Sec.II, we introduce our holographic setup. The effective string tension method with the anomalous flow to study confinement-deconfinement phase diagram is introduced in Sec. III. The entropy with chiral fluids is computed in Sec. IV. We make a short discussion and conclusion in Sec. V.

\section{The Setup}\label{sec:02}

To study the dependence of phase transition on anomalous flow, it is necessary to start
with the analysis of fluid/gravity duality \cite{Bhattacharyya:2007vjd,Erdmenger:2008rm,Banerjee:2008th,Megias:2013joa}. The action by using fluid/gravity duality $4 + 1$ dimensions \textrm{Einstein-Maxwell} theory \cite{Megias:2013joa} is
\begin{equation}
	\label{eq1}
S=\frac{1}{16\pi G}\int d^{5}x\sqrt{-g}[R+12-\frac{1}{4}F_{MN}F^{MN}+\frac{\kappa}{3}\epsilon^{MNPQR}A_{M}F_{NP}F_{QR}],
\end{equation}
where $G$ is Newton's constant, $F_{MN}$ and $A_{M}$ are a $4 + 1$- dimensional field strength and vector potential, and $\kappa$ is the $4 + 1$-dimensional Chern-Simons coupling, which is dual to the axial anomaly coefficient in the boundary gauge theory. The \textrm{Einstein-Maxwell} equations can be obtained from the action as
\begin{equation}
	\label{eq2}
G_{MN}+(\frac{1}{8}F^{2}-6)g_{MN}-\frac{1}{2}F_{ML}F_{N}^{L}=0,
\end{equation}
\begin{equation}
	\label{eq3}
\nabla_{N}F^{ND}+\epsilon^{DNPQR}\kappa F_{NP}F_{QR}=0,
\end{equation}
where the five-dimensional metric is selected to be of signature $(-, +, +, +, +)$, and the epsilon tensor $\epsilon_{ABCDE}=\sqrt{-g}\epsilon(ABCDE)$  has to be distinguished from the epsilon symbol $\epsilon(rtx^{1}x^{2}x^{3})=+1$.

The metric and gauge field discussing a static black hole with mass $m$ and charge $q$ are given as \cite{Megias:2013joa}
\begin{equation}
	\label{eq4}
ds^{2}=-r^{2}f(r)dt^{2}+\frac{1}{r^{2}f(r)dr^{2}}+r^{2}dx^{i}dx^{i},     A=\phi(r)dt,
\end{equation}
where
\begin{equation}
	\label{eq5}
f(r)=1-\frac{m}{r^{4}+\frac{q^{2}}{r^{6}}},   \phi(r)=-\frac{\sqrt{3}q}{2r^{2}}.
\end{equation}

$m$, $q$ are the mass parameter and charge parameter of the black hole respectively, which can be written as a function of temperature $T$ and chemical potential $\mu$ \cite{Son:2009tf} as
\begin{equation}
	\label{eq6}
m=\frac{\pi^{4}T^{4}}{16}(1+\sqrt{1+\frac{8}{3\pi^{2}}\overline{\mu}^{2}})^{3}(-1+3\sqrt{1+\frac{8}{3\pi^{2}}\overline{\mu}^{2}}),
\end{equation}
\begin{equation}
	\label{eq7}
q=\frac{\mu}{\sqrt{3}}\frac{\pi^{2}T^{2}}{2}(1+\sqrt{1+\frac{8}{3\pi^{2}}\overline{\mu}^{2}})^{2},
\end{equation}
where $\overline{\mu}=\frac{\mu}{T}$. By making $f(r)=0$, one obtains the two real roots as
\begin{equation}
	\label{eq8}
r_{+}=\frac{\pi T}{2}(1+\sqrt{1+\frac{8}{3\pi^{2}}\overline{\mu}^{2}}),
\end{equation}
\begin{equation}
	\label{eq9}
r_{-}=\sqrt{\frac{r_{+}^{2}}{2}(-1+\sqrt{9-\frac{8}{\frac{1}{2}(1+\sqrt{1+\frac{8}{3\pi^{2}}\overline{\mu}^{2}})}})},
\end{equation}
where $r_{+}$ is the outer horizon and $r_{-}$  is the inner horizon. In Eddington-Finkelstein coordinates, the metric and bulk gauge field to describe a static black hole with mass $m$ and charge $q$ take the form as
\begin{equation}
	\label{eq10}
ds^{2}=-r^{2}f(r)u_{\mu}u_{\nu}dx^{\mu}dx^{\nu}+r^{2}P_{\mu \nu}dx^{\mu}dx^{\nu}-2u_{\mu}dx^{\mu}dr,
\end{equation}
\begin{equation}
	\label{eq11}
A=-\phi(r)u_{\mu}dx^{\mu},
\end{equation}
where $u^{\mu}=(1,0,0,0)$, $P_{\mu \nu}=\eta_{\mu \nu}+u_{\mu}u_{\nu}$. The normalization condition is $u^{\mu}u_{\mu}=-1$.
It is found that Eqs.(\ref{eq10}) and (\ref{eq11}) is a solution of the equations of motion when  $m$, $q$ and $u^{\mu}$ are independent of the space-time coordinates $x^{\mu}$.
The fluid/gravity approach indicates that one has to promote all the parameters to slow varying functions of the space time coordinates, and contains corrections to the metric in order to turn it into a solution of the equations of motion again. By iteration one sets up the corrections proportional to first derivatives
\begin{equation}
	\label{eq12}
g_{AB}=g_{AB}^{(0)}+g_{AB}^{(1)}+...~~~~~ ,
\end{equation}
\begin{equation}
	\label{eq13}
A_{M}=A_{M}^{(0)}+A_{M}^{(1)}+...~~~~~~,
\end{equation}
requiring the solution to be regular at the horizon. For using the fluid/gravity techniques \cite{Bhattacharyya:2007vjd,Erdmenger:2008rm,Banerjee:2008th,Megias:2013joa}, one has to use a Weyl invariant formalism~\cite{Bhattacharyya:2008ji} as
\begin{equation}
\begin{split}
	\label{eq14}
ds^{2}=-2W_{1}(\rho)u_{\mu}dx^{\mu}(dr+r\textit{A}_{\nu}dx^{\nu})+r^{2}[W_{2}(\rho)\eta_{\mu \nu}+W_{3}(\rho)u_{\mu}u_{\nu}\\
+2\frac{W_{4}(\rho)}{r_{+}}P^{\sigma}_{\mu}u_{\nu}+\frac{W_{5}(\rho)}{r_{+}^{2}}]dx^{\mu}dx^{\nu},
\end{split}
\end{equation}
\begin{equation}
	\label{eq15}
A=(a_{\mu}^{(b)}+a_{\nu}(\rho)P^{\nu}_{\mu}+r_{+}c(\rho)u_{\mu})dx^{\mu},
\end{equation}
where $\rho=\frac{r}{r_{+}}$, $r_{+}=r_{+}(x^{\mu})$. The zero order solution of derivative expansion is
\begin{equation}
	\label{eq16}
c^{(0)}(\rho)=-\frac{\phi(\rho)}{r_{+}},
\end{equation}
\begin{equation}
	\label{eq17}
a_{\mu}^{(0)}(\rho)=0,
\end{equation}
\begin{equation}
	\label{eq18}
W_{1}^{(0)}(\rho)=W_{2}^{(0)}(\rho)=1,
\end{equation}
\begin{equation}
	\label{eq19}
W_{3}^{(0)}(\rho)=1-f(\rho),
\end{equation}
\begin{equation}
	\label{eq20}
W_{4}^{(0)}(\rho)=W_{5}^{(0)}(\rho)=0.
\end{equation}

By solving the following Einstein-Maxwell equation
\begin{equation}
	\label{eq21}
E_{\nu i}+r^{2}f(r)E_{ri}=0,
\end{equation}
\begin{equation}
	\label{eq22}
E_{ri}=0,
\end{equation}
\begin{equation}
	\label{eq23}
M_{i}=0,
\end{equation}
one can obtain the equation of order $n$ as
\begin{equation}
	\label{eq24}
\partial_{\rho}(\rho^{5}\partial_{\rho}W_{4i}^{(n)}+2\sqrt{3}Qa_{i}^{(n)}(\rho))=J_{i}^{(n)(\rho)},
\end{equation}
\begin{equation}
	\label{eq25}
\partial_{\rho}(\rho^{3}f(\rho)\partial_{\rho}a_{i}^{(n)}+2\sqrt{3}QW_{4i}^{(n)})=A_{i}^{(n)(\rho)},
\end{equation}

\noindent where $Q=\frac{q}{r_{+}^{3}}$ is Weyl invariable charge. In Ref.~\cite{Megias:2013joa}, the source terms $J_{i}^{(n)(\rho)}$  and $A_{i}^{(n)(\rho)}$  include the effects of vortex, electric field, magnetic field and gravitational anomalous term $\lambda$ , respectively. In this paper, the source term merely considers the magnetic field effect, and the first order sources are
\begin{equation}
	\label{eq26}
J_{\mu}^{(1)(\rho)}=0,
\end{equation}
\begin{equation}
	\label{eq27}
A_{\mu}^{(1)(\rho)}=-\frac{16\sqrt{3}Q\kappa B_{\mu}}{\rho^{3}r_{+}}.
\end{equation}

At this time, the first-order term in the metric (14) is simply
\begin{equation}
	\label{eq28}
W_{4\mu}^{(1)(\rho)}=F_{4}(\rho)\frac{B_{\mu}}{r_{+}},
\end{equation}
\begin{equation}
\begin{split}
	\label{eq29}
F_{4}(\rho)=\frac{\kappa(9Q^{4}\rho^{2}-3MQ^{2}\rho^{4}+6MQ^{2}\rho^{6})}{M\rho^{8}(1+2\rho_{2}^{2})^{3}}+\frac{6kQ^{2}f(\rho)log(\rho-\rho_{2})}{(1+2\rho_{2}^{2})^{3}}\\
+\frac{6kQ^{2}f(\rho)log(\rho+\rho_{2})}{(1+2\rho_{2}^{2})^{3}}+\frac{6kQ^{2}f(\rho)log(1+\rho^{2}+\rho_{2}^{2})}{(1+2\rho_{2}^{2})^{3}},
\end{split}
\end{equation}
where $\rho_{2}=\frac{r_{-}}{r_{+}}$,  and $M=\frac{m}{r_{+}^{4}}=1+Q^{2}$ Weyl invariable mass. The temperature and chemical potential can be reformulated as
\begin{equation}
	\label{eq30}
T=\frac{r_{+}}{2\pi}(2-Q^{2}),
\end{equation}
\begin{equation}
	\label{eq31}
\mu=\frac{\sqrt{3}r_{+}Q}{2}.
\end{equation}

If ${\mu}/{T}$ is very small, one can obtain
\begin{equation}
	\label{eq32}
\frac{W_{4\sigma}(\rho)}{r_{+}}=\frac{4\kappa \mu^{2}}{\pi^{4}T^{4}}Q(r)B_{\sigma},
\end{equation}
\begin{equation}
	\label{eq33}
Q(r)=\frac{2\pi^{2}T^{2}}{r^{2}}-\frac{\pi^{4}T^{4}}{r^{4}}+2(1-\frac{\pi^{4}T^{4}}{r^{4}})ln(\frac{r^{2}}{r^{2}+\pi^{2}T^{2}}).
\end{equation}

If the magnetic field is in the $x^{3}$ direction, it merely has $dx^{3}d\upsilon$ of the cross terms in the metric Eq.(\ref{eq14}) is not zero.
By using $dv=dt+\frac{1}{r^{2}f(r)dr}$, the cross term can be derived as
\begin{equation}
	\label{eq34}
2r^{2}\frac{W_{4\sigma}(\rho)}{r_{+}}P^{\sigma}_{\mu}u_{\nu}dx^{3}dv=-2r^{2}\frac{4\kappa \mu^{2}}{\pi^{4}T^{4}}Q(r)Bdx^{3}dt-2\frac{\frac{4\kappa \mu^{2}}{\pi^{4}T^{4}}Q(r)}{f(r)}Bdx^{3}dr.
\end{equation}

The pressure is $p=\frac{m}{16\pi G}$ \cite{Son:2009tf}, and the energy density is $\varepsilon=3p$ in a thermal \textrm{QGP} system. The term $\frac{4\kappa \mu^{2}}{\pi^{4}T^{4}}$ in Eq.{\ref{eq34}} can be written as
\begin{equation}
	\label{eq35}
\frac{4\kappa \mu^{2}}{\pi^{4}T^{4}}=\frac{4\kappa \mu^{2}}{\pi^{4}T^{4}\frac{m}{16\pi G}\frac{16\pi G}{m}}\approx \frac{\kappa \mu^{2}}{(\varepsilon+p)\pi G}.
\end{equation}

The entropy flow (in Appendix A) is calculated as
\begin{equation}
	\label{eq36}
     s^{\mu}=su^{\mu}-\frac{C\mu\mu_{A}}{\varepsilon+P}B^{\mu}=su^{\mu}+s\upsilon_{anom}.
\end{equation}

From $C=\frac{-k}{\pi G}$ \cite{Son:2009tf}, the anomalous flow related to the magnetic field can be derived as
\begin{equation}
	\label{eq37}
\frac{4\kappa \mu^{2}}{\pi^{4}T^{4}}B=-\frac{C\mu^{2}}{(2\varepsilon+p)}B=-\frac{C\mu \mu_{A}}{(\varepsilon+p)}B=\upsilon_{anom}.
\end{equation}

The detailed derivation process of anomalous flow $\upsilon_{anom}=-\frac{C\mu \mu_{A}}{(\varepsilon+p)}B$  is provided in Appendix A.  The metric which considers anomalous flow in the Landau frame of the fluid is given as
\begin{equation}
	\label{eq38}
ds^{2}=\frac{h(z)}{z^{2}}(-f(z)dt^{2}+d\vec{x}^{2}-2\upsilon_{\textrm{anom}}Q(z)dx_{3}dt)+\frac{2h(z)\upsilon_{\textrm{anom}}Q(z)}{z^{2}f(z)}dzdx_{3}+\frac{h(z)}{z^{2}f(z)}dz^{2},
\end{equation}
where $h(z)=e^{cz^{2}/2}$ is a wrap factor, which illustrates the characteristics of the soft wall model,
and the deformation parameter $c = 0.9 \textrm{GeV} ^{2}$ depicts the deviation from conformality \cite{Andreev:2006vy,Andreev:2006ct,Andreev:2010bv,Andreev:2006eh}.

\section{Effective string tension and phase diagram in the presence of anomalous flow }\label{sec:03}
By considering a $q\bar{q}$ dipole moving through a thermal plasma at the velocity $\upsilon=\tanh(\eta)$, we choose the plasma is at rest, and the dipole is moving with a constant rapidity $\eta$ along the $x_{3}$ direction. It is convenient to take the gravity background Eq.(\ref{eq38}) in the frame that the dipole is at rest while energy density of the \textrm{QGP} medium is moving with rapidity $-\eta$  in the $x_{3}$ direction. In our discussion, the $q\bar{q}$ dipole now is at rest and feels a hot plasma wind
\begin{equation}
	\label{eq39}
	dt=dt'\cosh(\eta)-dx_{3}'\sinh(\eta),
\end{equation}
\begin{equation}
	\label{eq40}
	dx_{3}=-dt'\sinh(\eta)+dx_{3}'\cosh(\eta).
\end{equation}

After dropping the prime, the metric is given as
\begin{equation}
\begin{split}
	\label{eq41}
ds^{2}=&\frac{h(z)}{z^{2}}[(-f(z)\cosh^{2}(\eta)+\sinh^{2}(\eta)+\upsilon_{\textrm{anom}}Q(z)\sinh(2\eta))dt^{2}+dx_{1}^{2}\\
       &+\frac{dz^{2}}{f(z)}+(-f(z)\sinh^{2}(\eta)+\cosh^{2}(\eta)+\upsilon_{\textrm{anom}}Q(z)\sinh(2\eta))dx_{3}^{2}\\
       &+(-\frac{2\upsilon_{\textrm{anom}}Q(z)}{f(z)}\sinh(\eta)dtdz+\frac{2\upsilon_{\textrm{anom}}Q(z)}{f(z)}\cosh(\eta)dx_{3}dz)]+dx_{2}^{2}\\
       &+(f(z)\sinh(2\eta)-\sinh(2\eta)-2\upsilon_{\textrm{anom}}Q(z)(\cosh^{2}(\eta)-\sinh^{2}(\eta)))dtdx_{3}.
\end{split}
\end{equation}

The Nambu-Goto action of the world sheet in the Minkowski metric is calculated as
\begin{equation}
	\label{eq42}
	S_{\textrm{NG}}=-\frac{1}{2\pi\alpha'}\int d^{2}\xi\sqrt{-\det g_{ab}},
\end{equation}

\noindent where $g_{ab}$ is the induced metric on the world sheet and $\frac{1}{2\pi\alpha'}$ is the string tension, and
\begin{equation}
	\label{eq43}
	g_{ab}=g_{MN}\partial_{a}X^{M}\partial_{b}X^{N},
\end{equation}

\noindent where $X^{M}$ and $g_{MN}$ are the coordinates and the metric of the AdS space. By using the static gauge $\xi^{0}=t$, $\xi^{1}=x_{2}$,
the Nambu-Goto action is given as
\begin{equation}
	\label{eq44}
S_{\textrm{NG}}=-\frac{1}{2\pi\alpha'T}\int_{-L/2}^{L/2}\textrm{d}x_{2}\sqrt{g_{2}(z)+g_{1}(z)(\frac{dz}{dx_{2}})^{2}
+f(z)(\frac{h(z)}{z^{2}}\frac{dx_{3}}{dx_{2}})^{2}+g_{3}(z)\frac{dx_{3}}{dx_{2}}\frac{dz}{dx_{2}}},
\end{equation}

\noindent where
\begin{equation}
	\label{eq45}
	g_{2}(z)=\frac{h^{2}(z)}{z^{4}}(f(z)\cosh^{2}(\eta)-\sinh^{2}(\eta)-\upsilon_{\textrm{anom}}Q(z)\sinh(2\eta)),
\end{equation}
\begin{equation}
	\label{eq46}
	g_{1}(z)=\frac{g_{2}(z)}{f(z)},    g_{3}(z)=2\upsilon_{\textrm{anom}}Q(z)\cosh(\eta)(\frac{h(z)}{z^{2}})^{2}.
\end{equation}

The separating distances of $q\bar{q}$ pair is given as
\begin{equation}
	\label{eq47}
	L=2\int_{0}^{z_{0}}[\frac{g_{2}(z)}{g_{1}(z)}(\frac{g_{2}(z)}{g_{2}(z_{0})}-1)]^{-1/2}dz.
\end{equation}

By considering the anomalous flow, the effective string tension is calculated as:
\begin{equation}
	\label{eq48}	\sigma(z)=\sqrt{g_{2}(z)}=\frac{h(z)}{z^{2}}\sqrt{f(z)\cosh^{2}(\eta)-\sinh^{2}(\eta)-\upsilon_{\textrm{anom}}Q(z)\sinh(2\eta)},
\end{equation}
where
\begin{equation}
	\label{eq49}	
\upsilon_{anom}\approx -0.003(\frac{eB}{T^{2}})(\frac{\mu_{V}\mu_{A}}{T^{2}})
\end{equation}
is the anomalous flow velocity, and its detailed derivation can refer to Appendix A. From Eq.{\ref{eq49}}, it can be observed that when $eB\rightarrow0$, the anomalous flow velocity $\upsilon_{anom}\rightarrow 0$ and there is no chiral magnetic effect. It can also be seen that even if $eB$ is large, the anomalous flow $\upsilon_{anom}$ is a small quantity. We often expand some physical quantities as $\upsilon_{anom}$ linear expansion at $\upsilon_{anom}=0$, taking only the contribution of the leading order in $\upsilon_{anom}$.

\begin{figure}
	\centering
	\includegraphics[width=8.5cm]{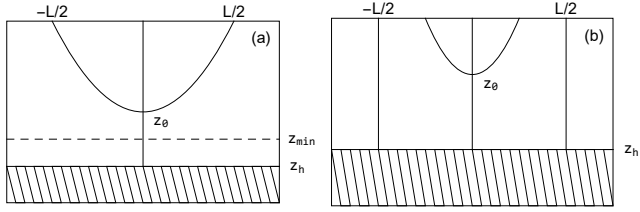}
	\caption{\label{fig1}(a) displays the characteristics of a U-shape open string, which links the $q\bar{q}$ with the confinement phase, and (b) displays two straight strings, which is at the separation distance when they reach the event horizon, corresponding to the deconfinement phase.}
\end{figure}

A fundamental string connects the $q\bar{q}$ as shown in Fig.~\ref{fig1}(a), and a U-shape open string connects the $q\bar{q}$ in the confinement phase. The deconfinement phase of the broken U-shape string is shown in Fig.~\ref{fig1}(b).

Figure~\ref{fig1}(a) shows a U-shape open string links the $q\bar{q}$ in the confinement phase. A dynamic wall $z_{\textrm{min}}$ exists at a low temperature, and the U-shape string can not exceed the dynamic wall with increasing the separating distances of the $q\bar{q}$ pair as shown in the situation of $T<T_{\textrm{C}}$ ($T_{\textrm{C}}$ ~phase transition temperature). $z_{0}$ is the lowest position of the U-shape string. When the temperature enhances to a certain value $(T_{\textrm{C}})$ and the value of horizon distance $z_{h}$ gets small, the system will experience a confinement-deconfinement phase transition. Figure~\ref{fig1}(b) shows the broken U-shape string, in which the broken string changes into two straight strings at separating distances, and reaches the position of the horizon. In the situation, a large black hole appears, and the dynamic wall disappears. The critical temperature $(T_{\textrm{C}})$ can be interpreted as a confinement-deconfinement phase transition temperature. The U-shape string still exists in the deconfinement phase if the distances $L$ of $q\bar{q}$ pair is small.

\begin{figure}
	\centering
	\includegraphics[width=8.5cm]{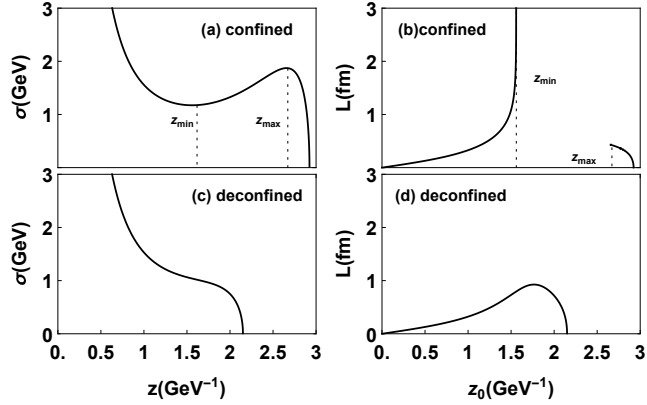}
	\caption{\label{fig2}The effective string tension corresponds to the confined situation (a) and
deconfined phase (c), respectively. The dependencies of interquark distances $L$ on the lowest position $z_{0}$ of the U-shape string are exhibited in the confined (b) and deconfined phase(d), respectively.}
\end{figure}

Equation~(\ref{eq48}) is the formula of effective string tension $\sigma(z)$ of the moving quarkonium with the anomalous flow. It is found that the effective string tension $\sigma(z)$ is a function of temperature, chemical potential, moving rapidity and anomalous flow $\upsilon_{\textrm{anom}}$. For fixed values of the chemical potential $\mu=0.1~ \textrm{GeV}$, and rapidity $\eta=0.3$ and temperature $T=0.1~\textrm{GeV}$, the dependence of the effective string tension $\sigma(z)$ on the fifth holographic coordinate $z$ in the confined phase is given in Fig. 2(a). The effective string tension $\sigma(z)$ reaches a minimum when $z=z_{\textrm{min}}$, and $\sigma(z)$ reaches the maximum value when $z=z_{\textrm{max}}$. The dependence of distance $L$ on $z_{0}$ in the confined phase is computed from the Eq.~(\ref{eq47}) as shown in Fig. \ref{fig2}(b). The distance increases monotonically from $L(z_{0})=0$ to $L(z_{\textrm{min}})\rightarrow \infty$  at  $0\leq z_{0} \leq z_{\textrm{min}}$. But at $z_{\textrm{max}}\leq z_{0} \leq z_{h}$ region, $L$ monotonically decreases from a finite value $L(z_\textrm{max})$ to $L(z_{h})=0$.

With the increase of temperature $T$ ($T<T_{\textrm{C}}$), the two points $z_{\textrm{min}}$ and $z_{\textrm{max}}$  get more and more close along the $z$ direction. When the temperature $T$ increases to a certain value such as a phase transition temperature $T_{\textrm{C}}$, the two points $z_{\textrm{min}}$ and $z_{\textrm{max}}$ coincide at $z_{\textrm{min}}=z_{\textrm{max}}=z_{m}$ . The effective string tension $\sigma(z)$ can be evaluated in Fig.~\ref{fig2}(c) by taking $\mu=0.1 ~\textrm{GeV}$ and $T=0.14~ \textrm{GeV}$. Figure~\ref{fig2}(d) shows that when $T>T_{C}$, $L(z)$ never exceeds the value $L(z_{m})$ at $0\leq z_{0} \leq z_{h}$.  For $L>L(z_{m})$ , two strings extend from the boundary $z_{0} = 0$ to the black hole horizon $z_{0} = z_{h}$ as shown in Fig.~\ref{fig1}(b). The quark-antiquark pair splits into two quarks.

From Eq.~(\ref{eq7}), the dependence of phase transition temperature $T_{\textrm{C}}$ on $q_{\textrm{C}}$ can be derived as
\begin{equation}
	\label{eq50}
     T_{\textrm{C}}=A q_{\textrm{C}}^{1/2}+\frac{B}{q_{\textrm{C}}^{1/2}},	
\end{equation}
\noindent where $A=\frac{\sqrt{27\sqrt{3}}}{3\pi\sqrt{6\mu}}$ and $B=-\frac{4\mu^{3}}{3\sqrt{3}}A$. By expanding phase transition temperature $T_{\textrm{C}}$ to the leading order at $\upsilon_{\textrm{anom}}=0$, the phase transition temperature $T_{\textrm{C}}$ can be given as
\begin{equation}
	\label{eq51}
     T_{\textrm{C}}=T_{\textrm{C}0}+\upsilon_{\textrm{anom}}(\frac{A}{2 q_{{\textrm{C}0}}^{1/2}}-\frac{B}{2 {q_{{\textrm{C}0}}^{3/2}}})q_{{\textrm{C}1}},
\end{equation}
\noindent where
\begin{equation}
	\label{eq52}
     T_{\textrm{C}0}=T_{\textrm{C}}(q_{{\textrm{C}0}})
\end{equation}
\noindent is the phase transition temperature when $\upsilon_{\textrm{anom}}=0$. Therefore the phase transition temperature $T_{\textrm{C}}$ can be written as
\begin{equation}
	\label{eq53}
     T_{\textrm{C}}=T_{\textrm{C}0}+\upsilon_{\textrm{anom}}T_{\textrm{C}1}.
\end{equation}

The resulting temperature change caused by anomalous flow is
\begin{equation}
	\label{eq54}
     T_{\textrm{C}1}=(\frac{A}{2 q_{{\textrm{C}0}}^{1/2}}-\frac{B}{2 {q_{{\textrm{C}0}}^{3/2}}})q_{{\textrm{C}1}},
\end{equation}

\noindent and the detailed derivation process of $q_{{\small C}}$, $q_{{\small C0}}$ and $q_{{\small C1}}$ are provided in the Appendix B.
\begin{figure}
	\centering
	\includegraphics[width=8.5cm]{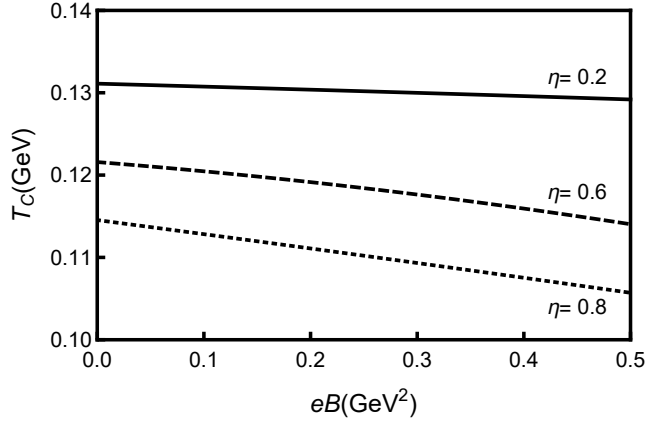}
	\caption{\label{fig3}The dependence of thermal AdS-black hole phase transition critical temperature
   $T_{\textrm{C}}$ on magnetic field $eB$ at various moving rapidity, where $\mu = 0.01~\textrm{GeV} $ is considered.}
\end{figure}
The dependence of phase transition critical temperature $T_{\textrm{C}}$ on $eB$ at various moving rapidity $\eta$ are shown in Fig.\ref{fig3}. It is found that the phase transition temperature $T_{\textrm{C}}$ decreases with the increase of magnetic field $eB$, which corresponds to the so-called inverse magnetic catalysis for the dual confinement-deconfinement transition. The conclusion is consistent with lattice \textrm{QCD }findings \cite{Bali:2012zg,Bali:2014kia}~and some \textrm{QCD} effective field theory \cite{Farias:2014eca,Ferreira:2014kpa,Ayala:2014iba,Mueller:2015fka}.  Figure~\ref{fig3} also shows that a moving system will reach the phase transition point at a lower temperature and chemical potential than a stationary system. The physical picture can be understood that a static quarkonium probe is placed in the expansively thermal \textrm{QGP} system, and the thermal \textrm{QGP} system moves through the probe with a certain rapidity. This means that the moving \textrm{QGP} thermal system needs to evolve for a long time from high temperature to a lower phase transition temperature, which correspondingly prolongs the lifetime of the \textrm{QGP} thermal system.
\begin{figure}
	\centering
	\includegraphics[width=8.5cm]{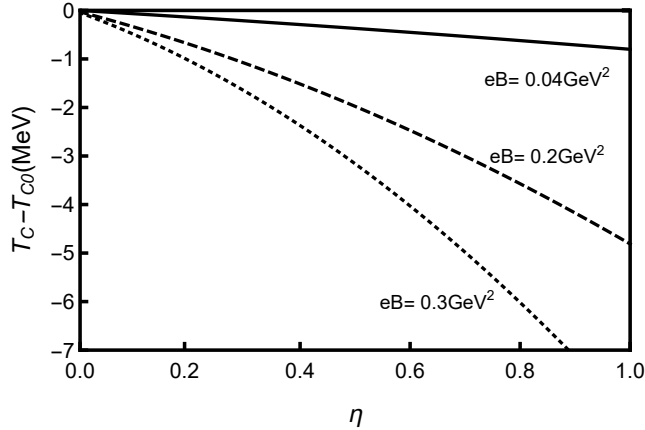}
	\caption{\label{fig4}The dependence of the change of phase transition temperature contributed by anomalous flow on moving rapidity $\eta$ at various magnetic fields ($eB$).}
\end{figure}

The dependence of the change of phase transition temperature contributed by anomalous flow on moving rapidity $\eta$ at various magnetic field $eB$ is shown in Fig.~\ref{fig4}. Under the same dipole moving rapidity, the greater the magnetic field $eB$ is, the greater the reduction of phase transition temperature is. On the other hand, under the same magnetic field $eB$, the faster the moving dipole is, the greater the reduction of phase transition temperature is.

\section{The Free energy and entropy of a moving quarkonium in a background magnetic field}\label{sec:04}
The free energy of $q\bar{q}$  pair is taken as:
\begin{equation}
	\label{eq55}
    \frac{\pi}{\sqrt{\lambda}}F_{q\bar{q}}=\int_{0}^{z_{0}}(\sqrt{\frac{g_{2}(z)g_{1}(z)}{g_{2}(z)-g_{20}(z_{0}^{(0)})}}-\sqrt{g_{2}(z\rightarrow 0)})dz-\int_{z_{0}}^{\infty}\sqrt{g_{2}(z\rightarrow 0)}dz,
\end{equation}
by expanding $z_{0}$ to the leading order in $\upsilon_{\textrm{anom}}$ in the presence of the anomalous flow, one can obtain
\begin{equation}
	\label{eq56}
    z_{0} \approx z_{0}^{(0)}+\upsilon_{\textrm{anom}}z_{1}+\emph{O}(\upsilon_{\textrm{anom}}^{2}).
\end{equation}
Let $\tilde{z}=\frac{z}{z_{0}}$, and the free energy is shown as
\begin{equation}
	\label{eq57}
     \frac{\pi}{\sqrt{\lambda}}F_{q\bar{q}}=\int_{0}^{1}(\sqrt{\frac{g_{2}(\tilde{z}z_{0})g_{1}(\tilde{z}z_{0})}{g_{2}(\tilde{z}z_{0})-g_{20}(z_{0}^{(0)})}}-\sqrt{g_{2}(\tilde{z}z_{0}\rightarrow 0)})d\tilde{z}z_{0}-\int_{1}^{\infty}\sqrt{g_{2}(\tilde{z}z_{0}\rightarrow 0)}d\tilde{z}z_{0},
\end{equation}
where
\begin{equation}
	\label{eq58}
   g_{2}(\tilde{z}z_{0})\approx g_{20}(\tilde{z}z_{0}^{(0)})+(g_{20}^{'}(\tilde{z}z_{0}^{(0)})\tilde{z}z_{1}+g_{21}(\tilde{z}z_{0}^{(0)}))\upsilon_{\textrm{anom}},
\end{equation}
\begin{equation}
	\label{eq59}
     g_{1}(\tilde{z}z_{0}) \approx g_{10}(\tilde{z}z_{0}^{(0)})+(g_{10}^{'}(\tilde{z}z_{0}^{(0)})\tilde{z}z_{1}+g_{11}(\tilde{z}z_{0}^{(0)}))\upsilon_{\textrm{anom}}.
\end{equation}
Let $z^{'}=\tilde{z}z_{0}^{(0)}$, Eq.(\ref{eq58}) and Eq.(\ref{eq59}) become to
\begin{equation}
	\label{eq60}
      g_{2}(z'\frac{z_{0}}{z_{0}^{(0)}}) \approx g_{20}(z')+(g'_{20}(z')\frac{z'z_{1}}{z^{(0)}}+g_{21}(z'))\upsilon_{\textrm{anom}},
\end{equation}
\begin{equation}
	\label{eq61}
      g_{1}(z'\frac{z_{0}}{z_{0}^{(0)}}) \approx g_{10}(z')+(g'_{10}(z')\frac{z'z_{1}}{z^{(0)}}+g_{11}(z'))\upsilon_{\textrm{anom}}.
\end{equation}
After dropping the prime, one obtains the free energy as
\begin{equation}
	\label{eq62}
       \frac{\pi}{\sqrt{\lambda}}F_{q\bar{q}} = \frac{z_{0}}{z_{0}^{(0)}}\int_{0}^{z_{0}^{(0)}}(\sqrt{\frac{g_{2}(z\frac{z_{0}}{z_{0}^{(0)}})g_{1}(z\frac{z_{0}}{z_{0}^{(0)}})}{g_{2}(z\frac{z_{0}}{z_{0}^{(0)}})-g_{20}(z_{0}^{(0)})}}-\sqrt{g_{2}(z\rightarrow 0)})dz-\frac{z_{0}}{z_{0}^{(0)}}\int_{z_{0}^{(0)}}^{\infty}\sqrt{g_{2}(z\rightarrow 0)}dz,
\end{equation}
where $\lambda=2.6$. When $\upsilon_{\textrm{anom}}=0$, then free energy of $q\bar{q}$ is calculated as
\begin{equation}
	\label{eq63}
       \frac{\pi}{\sqrt{\lambda}}F_{q\bar{q}}=\int_{0}^{z_{0}^{(0)}}(\sqrt{\frac{g_{20}(z)g_{10}(z)}{g_{20}(z)-g_{20}(z_{0}^{(0)})}}-\sqrt{g_{20}(z\rightarrow 0)})dz-\int_{z_{0}^{(0)}}^{\infty}\sqrt{g_{20}(z\rightarrow 0)}dz.
\end{equation}
The entropy of $q\bar{q}$ pair can be given as
\begin{equation}
	\label{eq64}
      S_{q\bar{q}}=-\frac{\partial F_{q\bar{q}}}{\partial T},
\end{equation}
\noindent when $\upsilon_{\textrm{anom}}=0$, the free energy of $q\bar{q}$ is written as
\begin{equation}
	\label{eq65}
      S_{q\bar{q}}^{(0)}=-\frac{\partial F_{q\bar{q}}^{(0)}}{\partial T}.
\end{equation}

\begin{figure}
	\centering
	\includegraphics[width=8.5cm]{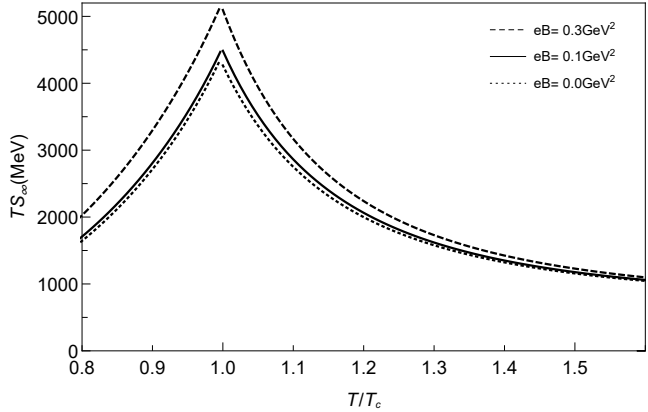}
	\caption{\label{fig5}The dependencies of asymptotic entropy ($TS_{\infty}$) on temperature ($T/T_{\textrm{C}}$) at different magnetic field.}
\end{figure}

The calculated entropy distribution in Fig.\ref{fig5} is the asymptotic entropy distribution $TS_{\infty}(T,r\rightarrow \infty)=-T\frac{\partial F_{\infty}}{\partial T}$, $r\rightarrow \infty$ means that $r$ reaches the maximum, and quarkonium becomes independent quark and antiquark. This distance between quark and antiquark is called the separation distance at this time. When the quark-antiquark distance is very small, the free energy has no or insignificant medium effect. When the quark-antiquark distance is close to the separation distance, the free energy $F(T)$  is merely a function of temperature, and the influence of the medium is significant. The separation distance of the heavy quarkonium is fixed, so the entropy distribution $TS_{\infty}$ of the quarkonium increases with temperature in the confined phase. In the deconfined phase, the separation distance of quarkonium decreases with temperature, and the entropy distribution $TS_{\infty}$ decreases with temperature. The features of entropy distribution $TS_{\infty}$ computed by us are consistent with lattice results \cite{Hashimoto:2014fha,Kaczmarek:2002mc,Petreczky:2004pz,Kaczmarek:2005zp}.
\begin{figure}
	\centering
	\includegraphics[width=15.0cm]{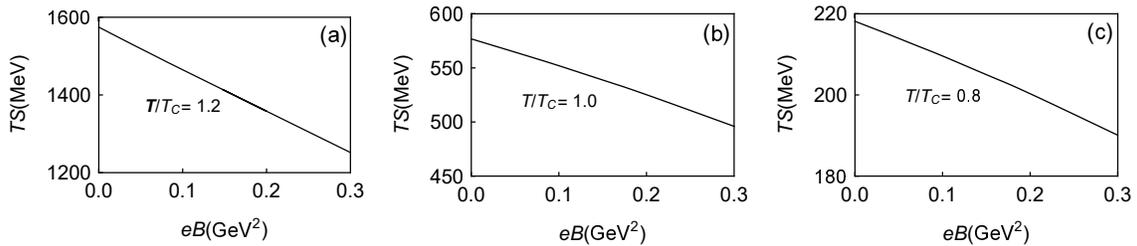}
	\caption{\label{fig6}The dependencies of entropy ($TS$) on magnetic field ($eB$) in the deconfinement phase ($T/T_{\textrm{C}}=1.2$) (a), Confinement-deconfinement phase transition ($T/T_{\textrm{C}}=1$) (b) and confinement phase ($T/T_{\textrm{C}}=0.8$)(c). We fixed the quark antiquark distance at 0.6 \textrm{fm}.}
\end{figure}

We fixed the quark antiquark distance at 0.6 \textrm{fm} in Fig.\ref{fig6}. The dependencies of entropy ($TS$) on magnetic field ($eB$) in the deconfinement phase ($T/T_{\textrm{C}}=1.2$), confinement-deconfinement phase transition ($T/T_{\textrm{C}}=1$) and confinement phase ($T/T_{\textrm{C}}=0.8$) are shown in Fig. 6(a,b,c), respectively. The entropy ($TS$) is a monotonic decreasing function of magnetic field in these three different cases. The maximum is close to 1600~\textrm{MeV} in the deconfinement phase with $T/T_{\textrm{C}}=1.2$, and close to 220~\textrm{MeV} in the confinement phase $T/T_{\textrm{C}}=0.8$. We computer the dependencies of entropy ($TS$) on magnetic field ($eB$) in the confinement-deconfinement phase transition $T/T_{\textrm{C}}=1$.

\section{Summary and Conclusions}\label{sec:05}
It is generally believed that there will be a large magnetic field, and the confinement-deconfinement phase transition also occurs in the early stage of relativistic heavy ion noncentral collision in the \textrm{RHIC} and \textrm{LHC} energy region. The strong magnetic field will produce a novel anomalous transport phenomenon, which corresponds to the chiral magnetic effect and induces anomalous flow.  The concept of anomalous flow is introduced to study the characteristics of the confinement-deconfinement phase transition of RHIC and LHC energy region: in the range of magnetic field (less than $0.3~ \textrm{GeV}^{2}$) with small chemical potential.

The fluid/gravity correspondence is a very powerful tool to understand the hydrodynamic dynamics of quantum field theories with holographic dual. The application of fluid/gravity correspondence to theories including chiral anomalies to study phase transition with chiral magnetic effects. In this paper, we assume that the initial hydrodynamic environment is a \textrm{QGP} phase where merely u and d quarks are considered when adding a magnetic field. The quark gluon plasma produced by relativistic heavy ion collision will produce anomalous flow. The chiral magnetic current formed by these two quarks has an impact on the heavy quarkonium.

We investigate the deconfinement phase transition by analyzing the characteristics of the effective string tension with anomalous flow in a hot plasma wind to gain insight into the influence such a field can have on crucial \textrm{QCD} observable. The dependence of phase transition temperature $T_{C}$ on $eB$ at different moving rapidity $\eta$ are provided, which is also one of the main results of this paper. In particular, $T_{C}$ decreases with the magnetic field caused by anomalous flow. The deconfined phase transition manifests the characteristics of inverse magnetic catalysis, which is in qualitatively agreement with the lattice \textrm{QCD} findings and some \textrm{QCD} effective field theory. From our study, a moving system will reach the phase transition point at a lower temperature and a smaller chemical potential than a stationary system. The physical picture can be understood that a static quarkonium probe is placed in the expansive thermal \textrm{QGP} system, and the thermal \textrm{QGP} system moves through the probe with a certain rapidity. This means that the moving \textrm{QGP} thermal system needs to evolve for a long time to arrive at a lower phase transition temperature, which correspondingly prolongs the lifetime of the moving \textrm{QGP} thermal system.

The dependencies of entropy ($TS$) on magnetic field ($eB$) in the deconfinement phase, confinement-deconfinement phase transition and confinement phase are computed , respectively. It is found that the entropy ($TS$) is a monotonic decreasing function of magnetic field in these three different cases.

\section*{Acknowledgments}
This work was supported by National Natural Science Foundation of China (Grants No. 11875178, No. 11475068, No. 11747115).
\appendix
\section{Calculation of anomalous flow velocity}
\begin{verbatim}
\end{verbatim}
\vskip-1.5cm
Our anomalous flow velocity formula comes from Son and Surowka’s article \cite{Son:2009tf}, in which they considered magnetic field and vortex effects, while we merely take into account the effect of magnetic field in the paper. The specific derivation process is given as follows. The entropy current is given as \cite{Son:2009tf}
\begin{equation}
	\label{eqA1}
     s^{\mu}=su^{\mu}-\frac{\mu}{T}\xi_{B}B^{\mu}+D_{B}B^{\mu},
\end{equation}
where $\xi_{B}=C(\mu-\frac{1}{2}\frac{n\mu^{2}}{\varepsilon+p})$ is the dissipation coefficient caused by magnetic field related to the dissipation of particle current, and $D_{B}=\frac{C\mu^{2}}{2T}$ is dissipation coefficient caused by magnetic field related to the dissipation of entropy current. By inserting the dissipation coefficients into formula (\ref{eqA1}), the entropy current $s^{\mu}$ can be written as
\begin{equation}
	\label{eqA2}
     s^{\mu}=su^{\mu}-\frac{\mu}{T}C(\mu-\frac{1}{2}\frac{n\mu^{2}}{\varepsilon+p})B^{\mu}+\frac{C\mu^{2}}{2T}B^{\mu}
            =su^{\mu}-\frac{C\mu^{2}}{2T}(1-\frac{n\mu}{\varepsilon+p})B^{\mu}.
\end{equation}

From thermodynamic relation $\varepsilon+p=Ts+\mu n$, the entropy current is given as
\begin{equation}
	\label{eqA3}
     s^{\mu}=su^{\mu}-\frac{C\mu^{2}}{2T}(1-\frac{\varepsilon+p-Ts}{\varepsilon+p})B^{\mu}=su^{\mu}-s\frac{C\mu^{2}}{2\varepsilon+p}B^{\mu}.
\end{equation}

When only right-handed fermions are considered, the entropy current is calculated as:
\begin{equation}
	\label{eqA4}
     s^{\mu}=su^{\mu}-\frac{C\mu\mu_{A}}{\varepsilon+P}B^{\mu}=su^{\mu}+s\upsilon_{anom}.
\end{equation}

From the dissipative part of entropy current $s\upsilon_{anom}$ , the anomalous flow velocity can be written as
\begin{equation}
	\label{eqA5}
     \upsilon_{anom}=-\frac{C\mu\mu_{A}}{\varepsilon+p}B^{\mu}.
\end{equation}

The parameter $C$ in Eq.(\ref{eqA5}) is given as
\begin{equation}
	\label{eqA6}
    C=\frac{C_{EM}e^{2}}{2\pi^{2}}.
\end{equation}

If only $u$ and $d$ quarks are considered, the contributions to \textrm{CME} are given with $N_{f}=2$, $C_{EM} = 5/9$.  The equations of state of a massless ideal quark-gluon gas are given as
\begin{equation}
	\label{eqA7}
     p=\frac{g_{QGP}\pi^{2}}{90}T^{4}, \hskip1.2cm   \varepsilon=3p,
\end{equation}
where $g_{QGP}=g_{G}+\frac{7}{8}g_{Q}$ is the number of degrees of freedom with gluon $g_{G}=(N_{C}^{2}-1)N_{S}$ and with quark $g_{Q}=N_{f}N_{C}N_{S}$. $N_{C}=3$, $N_{f}=2$ and $N_{S}=2$ are the numbers of colors, flavors and spin states, respectively. From Eq.(\ref{eqA5}), Eq.(\ref{eqA6}) and Eq.(\ref{eqA7}), the anomalous flow velocity can be obtained as
\begin{equation}
	\label{eqA8}
    \upsilon_{anom}\approx -0.003(\frac{eB}{T^{2}})(\frac{\mu_{V}\mu_{A}}{T^{2}}).
\end{equation}

\section{The dependence of $q_{\textrm{c}}$ and $q_{\textrm{c}0}$ in the presence of anomalous flow during phase transition}
\begin{verbatim}
\end{verbatim}
\vskip-1.5cm

When $\upsilon_{\textrm{anom}}\neq 0$, one should make $d\sigma/dz=\dot{\sigma}=0$ and $d^{2}\sigma/dz^{2}=\ddot{\sigma}=0$  to study the phase translation, and the corresponding formulas can be written as
\begin{equation}
	\label{eqB1}
      (-2cMz^{6}+2cq^{2}z^{8}+2q^{2}z^{6})\cosh^{2}\eta+2cz^{2}-4-\upsilon_{\textrm{anom}}\sinh2\eta[(2cz^{2}-4)Q+\dot{Q}z]=0 ,
\end{equation}
and
\begin{small}
\begin{equation}
	\label{eqB2}
      (-12cMz^{5}+16cq^{2}z^{7}+12q^{2}z^{5})\cosh^{2}\eta+4cz-\upsilon_{\textrm{anom}}\sinh2\eta[4czQ+(2cz^{2}-3)\dot{Q}+\ddot{Q}z]=0,
\end{equation}
\end{small}
where $\dot{Q}=\frac{\textrm{d}Q}{\textrm{d}z}$, $\ddot{Q}=\frac{\textrm{d}^{2}Q}{\textrm{d}z^{2}}$.

By solving Eqs.(\ref{eqB1}) and Eqs.(\ref{eqB2}) simultaneously, the $q_{C}^{2}$ formula at phase transition $T=T_{\textrm{C}}$, $z=z_{\textrm{c}}$ in the presence of anomalous flow can be written as
\begin{equation}
	\label{eqB3}
      q_{\textrm{C}}^{2}=\frac{2cz_{c}^{2}-6}{cz_{c}^{8}\cosh^{2}\eta}+\upsilon_{\textrm{anom}}\frac{\sinh2\eta}{cz_{c}^{6}
      \cosh^{2}\eta}(\frac{-2cz_{c}^{2}+6}{z_{c}^{2}}Q-\frac{9-2z_{c}^{2}}{4z_{c}}\dot{Q}+\frac{1}{4}\ddot{Q}).
\end{equation}

When $\upsilon_{\textrm{anom}}=0$, which corresponds to $T=T_{\textrm{C}0}$, the related result of $q_{\textrm{C}}$ is given as
\begin{equation}
	\label{eqB4}
      q_{\textrm{C}0}=\sqrt{\frac{2cz_{c0}^{2}-6}{cz_{c}^{8}\cosh^{2}\eta}}.
\end{equation}

In the presence of the anomalous flow, $z=z_{\textrm{c}}$ can be expanded to the leading order in $\upsilon_{\textrm{anom}}$ as
\begin{equation}
	\label{eqB5}
      z_{c}\approx z_{c0}+\upsilon_{\textrm{anom}}z_{c1}+ \emph{O}(\upsilon^{2}_{\textrm{anom}}).
\end{equation}

$g_{2}(z_{c})$  can be expanded by linear perturbation at $\upsilon_{\textrm{anom}}=0$ as
\begin{equation}
	\label{eqB6}
      g_{2}(z_{c})=g_{20}(z_{c})+g_{21}(z_{c})\upsilon_{\textrm{anom}}+ \emph{O}(\upsilon^{2}_{\textrm{anom}}),
\end{equation}

\noindent from Eqs.(\ref{eq45}) and Eqs.(\ref{eqB6}), $g_{21}(z_{\textrm{c}})$ can be derived as
\begin{equation}
	\label{eqB7}
     g_{21}(z_{c})=-\frac{h^{2}(z_{c})}{z_{c}^{4}}Q(z_{c})\sinh2\eta,
\end{equation}
\begin{equation}
	\label{eqB8}
     g_{20}(z_{c})=\frac{h^{2}(z_{c})}{z_{c}^{4}}(f(z_{c})\cosh^{2}\eta-\sinh^{2}\eta).
\end{equation}

By expanding $g_{20}(z_{c})$ at $z_{c}=z_{c0}$ ,  one can obtain
\begin{equation}
\begin{split}
	\label{B9}
g_{2}(z_{c})&=g_{20}(z_{c})+g_{21}(z_{c0})\upsilon_{\textrm{anom}}\\
    &=g_{20}(z_{c0})+g'_{20}(z_{c0})(z_{c}-z_{c0})+g_{21}(z_{c0})\upsilon_{\textrm{anom}}\\
    &=g_{20}(z_{c0})+g'_{20}(z_{c0})\upsilon_{\textrm{anom}}z_{c1}+g_{21}(z_{c0})\upsilon_{\textrm{anom}}\\
    &=g_{20}(z_{c0}),
\end{split}
\end{equation}

\noindent and
\begin{equation}
	\label{eqB10}
     z_{c1}=-\frac{g_{21}(z_{\textrm{c}0})}{g'_{20}(z_{\textrm{c}0})}.
\end{equation}

Substituting formulas (\ref{eqB7}), (\ref{eqB8}), and (\ref{eqB10}) into formula (\ref{eqB3}), one can get the result as
\begin{equation}
	\label{eqB11}
    q_{\textrm{C}}^{2}=\frac{2cz_{c0}^{2}-6}{cz_{c0}^{8}\cosh^{2}\eta}+\upsilon_{\textrm{anom}}G(\eta,Q,\dot{Q},\ddot{Q}),
\end{equation}

\noindent where
\begin{equation}
	\label{eqB12}
    G(\eta,Q,\dot{Q},\ddot{Q})=\frac{\sinh2\eta}{cz_{c0}^{6}\cosh^{2}\eta}(\frac{48-12cz_{c0}^{2}}{z_{c0}^{3}\sinh2\eta}
    +\frac{-2c0z_{c0}^{2}+6}{z_{c0}^{2}}Q-\frac{9-2z_{c0}^{2}}{4z_{c0}}\dot{Q}+\frac{1}{4}\ddot{Q}),
\end{equation}

\noindent we can get
\begin{equation}
	\label{eqB13}
     q_{\textrm{C}}^{2}=q_{\textrm{C}0}^{2}+\upsilon_{\textrm{anom}}G(\eta,Q,\dot{Q},\ddot{Q}).
\end{equation}

By expanding $q_{\textrm{C}}$  to the leading order in $\upsilon_{\textrm{anom}}$ in the presence of the anomalous flow, one can obtain
\begin{equation}
	\label{eqB14}
     q_{\textrm{C}}=q_{\textrm{C}0}+\upsilon_{anom}q_{\textrm{C}1}+O(\upsilon^{2}_{\textrm{anom}}).
\end{equation}

By squaring both sides of Eq.(\ref{eqB14}) simultaneously and ignoring the higher-order term of $\upsilon^{2}_{\textrm{anom}}$ , one
can obtain
\begin{equation}
	\label{eqB15}
     q_{\textrm{C}}^{2}=q_{\textrm{C}0}^{2}+2\upsilon_{\textrm{anom}}q_{\textrm{C}0}q_{\textrm{C}1}.
\end{equation}

From Eq.(\ref{eqB13}) and Eq.(\ref{eqB15}), $q_{\textrm{C}1}$  can be given as
\begin{equation}
	\label{eqB16}
     q_{\textrm{C}1}=\frac{G(\eta,Q,\dot{Q},\ddot{Q})}{2q_{\textrm{C}0}},
\end{equation}

\noindent where $q_{\textrm{C}0}$ is given by Eq.(\ref{eqB4}).

\section*{References}

\bibliography{ref}

\begin{thebibliography}{53}%
\makeatletter
\providecommand \@ifxundefined [1]{%
 \@ifx{#1\undefined}
}%
\providecommand \@ifnum [1]{%
 \ifnum #1\expandafter \@firstoftwo
 \else \expandafter \@secondoftwo
 \fi
}%
\providecommand \@ifx [1]{%
 \ifx #1\expandafter \@firstoftwo
 \else \expandafter \@secondoftwo
 \fi
}%
\providecommand \natexlab [1]{#1}%
\providecommand \enquote  [1]{``#1''}%
\providecommand \bibnamefont  [1]{#1}%
\providecommand \bibfnamefont [1]{#1}%
\providecommand \citenamefont [1]{#1}%
\providecommand \href@noop [0]{\@secondoftwo}%
\providecommand \href [0]{\begingroup \@sanitize@url \@href}%
\providecommand \@href[1]{\@@startlink{#1}\@@href}%
\providecommand \@@href[1]{\endgroup#1\@@endlink}%
\providecommand \@sanitize@url [0]{\catcode `\\12\catcode `\$12\catcode
  `\&12\catcode `\#12\catcode `\^12\catcode `\_12\catcode `\%12\relax}%
\providecommand \@@startlink[1]{}%
\providecommand \@@endlink[0]{}%
\providecommand \url  [0]{\begingroup\@sanitize@url \@url }%
\providecommand \@url [1]{\endgroup\@href {#1}{\urlprefix }}%
\providecommand \urlprefix  [0]{URL }%
\providecommand \Eprint [0]{\href }%
\providecommand \doibase [0]{http://dx.doi.org/}%
\providecommand \selectlanguage [0]{\@gobble}%
\providecommand \bibinfo  [0]{\@secondoftwo}%
\providecommand \bibfield  [0]{\@secondoftwo}%
\providecommand \translation [1]{[#1]}%
\providecommand \BibitemOpen [0]{}%
\providecommand \bibitemStop [0]{}%
\providecommand \bibitemNoStop [0]{.\EOS\space}%
\providecommand \EOS [0]{\spacefactor3000\relax}%
\providecommand \BibitemShut  [1]{\csname bibitem#1\endcsname}%
\let\auto@bib@innerbib\@empty
\bibitem [{\citenamefont {Matsui}\ and\ \citenamefont
  {Satz}(1986)}]{Matsui:1986dk}%
  \BibitemOpen
  \bibfield  {author} {\bibinfo {author} {\bibfnamefont {T.}~\bibnamefont
  {Matsui}}\ and\ \bibinfo {author} {\bibfnamefont {H.}~\bibnamefont {Satz}},\
  }\href {\doibase 10.1016/0370-2693(86)91404-8} {\bibfield  {journal}
  {\bibinfo  {journal} {Phys. Lett. B}\ }\textbf {\bibinfo {volume} {178}},\
  \bibinfo {pages} {416} (\bibinfo {year} {1986})}\BibitemShut {NoStop}%
\bibitem [{\citenamefont {Rapp}\ \emph {et~al.}(2010)\citenamefont {Rapp},
  \citenamefont {Blaschke},\ and\ \citenamefont {Crochet}}]{Rapp:2008tf}%
  \BibitemOpen
  \bibfield  {author} {\bibinfo {author} {\bibfnamefont {R.}~\bibnamefont
  {Rapp}}, \bibinfo {author} {\bibfnamefont {D.}~\bibnamefont {Blaschke}}, \
  and\ \bibinfo {author} {\bibfnamefont {P.}~\bibnamefont {Crochet}},\ }\href
  {\doibase 10.1016/j.ppnp.2010.07.002} {\bibfield  {journal} {\bibinfo
  {journal} {Prog. Part. Nucl. Phys.}\ }\textbf {\bibinfo {volume} {65}},\
  \bibinfo {pages} {209} (\bibinfo {year} {2010})},\ \Eprint
  {http://arxiv.org/abs/0807.2470} {arXiv:0807.2470 [hep-ph]} \BibitemShut
  {NoStop}%
\bibitem [{\citenamefont {Satz}(2006)}]{Satz:2005hx}%
  \BibitemOpen
  \bibfield  {author} {\bibinfo {author} {\bibfnamefont {H.}~\bibnamefont
  {Satz}},\ }\href {\doibase 10.1088/0954-3899/32/3/R01} {\bibfield  {journal}
  {\bibinfo  {journal} {J. Phys. G}\ }\textbf {\bibinfo {volume} {32}},\
  \bibinfo {pages} {R25} (\bibinfo {year} {2006})},\ \Eprint
  {http://arxiv.org/abs/hep-ph/0512217} {arXiv:hep-ph/0512217} \BibitemShut
  {NoStop}%
\bibitem [{\citenamefont {Liu}\ \emph {et~al.}(2007)\citenamefont {Liu},
  \citenamefont {Rajagopal},\ and\ \citenamefont {Wiedemann}}]{Liu:2006nn}%
  \BibitemOpen
  \bibfield  {author} {\bibinfo {author} {\bibfnamefont {H.}~\bibnamefont
  {Liu}}, \bibinfo {author} {\bibfnamefont {K.}~\bibnamefont {Rajagopal}}, \
  and\ \bibinfo {author} {\bibfnamefont {U.~A.}\ \bibnamefont {Wiedemann}},\
  }\href {\doibase 10.1103/PhysRevLett.98.182301} {\bibfield  {journal}
  {\bibinfo  {journal} {Phys. Rev. Lett.}\ }\textbf {\bibinfo {volume} {98}},\
  \bibinfo {pages} {182301} (\bibinfo {year} {2007})},\ \Eprint
  {http://arxiv.org/abs/hep-ph/0607062} {arXiv:hep-ph/0607062} \BibitemShut
  {NoStop}%
\bibitem [{\citenamefont {Feng}\ \emph {et~al.}(2020)\citenamefont {Feng},
  \citenamefont {Zhao},\ and\ \citenamefont {Chen}}]{Feng:2019boe}%
  \BibitemOpen
  \bibfield  {author} {\bibinfo {author} {\bibfnamefont {S.-Q.}\ \bibnamefont
  {Feng}}, \bibinfo {author} {\bibfnamefont {Y.-Q.}\ \bibnamefont {Zhao}}, \
  and\ \bibinfo {author} {\bibfnamefont {X.}~\bibnamefont {Chen}},\ }\href
  {\doibase 10.1103/PhysRevD.101.026023} {\bibfield  {journal} {\bibinfo
  {journal} {Phys. Rev. D}\ }\textbf {\bibinfo {volume} {101}},\ \bibinfo
  {pages} {026023} (\bibinfo {year} {2020})},\ \Eprint
  {http://arxiv.org/abs/1910.05668} {arXiv:1910.05668 [hep-ph]} \BibitemShut
  {NoStop}%
\bibitem [{\citenamefont {Sadofyev}\ and\ \citenamefont
  {Yin}(2016)}]{Sadofyev:2015hxa}%
  \BibitemOpen
  \bibfield  {author} {\bibinfo {author} {\bibfnamefont {A.~V.}\ \bibnamefont
  {Sadofyev}}\ and\ \bibinfo {author} {\bibfnamefont {Y.}~\bibnamefont {Yin}},\
  }\href {\doibase 10.1007/JHEP01(2016)052} {\bibfield  {journal} {\bibinfo
  {journal} {JHEP}\ }\textbf {\bibinfo {volume} {01}},\ \bibinfo {pages} {052}
  (\bibinfo {year} {2016})},\ \Eprint {http://arxiv.org/abs/1510.06760}
  {arXiv:1510.06760 [hep-th]} \BibitemShut {NoStop}%
\bibitem [{\citenamefont {Skokov}\ \emph {et~al.}(2009)\citenamefont {Skokov},
  \citenamefont {Illarionov},\ and\ \citenamefont {Toneev}}]{Skokov:2009qp}%
  \BibitemOpen
  \bibfield  {author} {\bibinfo {author} {\bibfnamefont {V.}~\bibnamefont
  {Skokov}}, \bibinfo {author} {\bibfnamefont {A.~Y.}\ \bibnamefont
  {Illarionov}}, \ and\ \bibinfo {author} {\bibfnamefont {V.}~\bibnamefont
  {Toneev}},\ }\href {\doibase 10.1142/S0217751X09047570} {\bibfield  {journal}
  {\bibinfo  {journal} {Int. J. Mod. Phys. A}\ }\textbf {\bibinfo {volume}
  {24}},\ \bibinfo {pages} {5925} (\bibinfo {year} {2009})},\ \Eprint
  {http://arxiv.org/abs/0907.1396} {arXiv:0907.1396 [nucl-th]} \BibitemShut
  {NoStop}%
\bibitem [{\citenamefont {Voronyuk}\ \emph {et~al.}(2011)\citenamefont
  {Voronyuk}, \citenamefont {Toneev}, \citenamefont {Cassing}, \citenamefont
  {Bratkovskaya}, \citenamefont {Konchakovski},\ and\ \citenamefont
  {Voloshin}}]{Voronyuk:2011jd}%
  \BibitemOpen
  \bibfield  {author} {\bibinfo {author} {\bibfnamefont {V.}~\bibnamefont
  {Voronyuk}}, \bibinfo {author} {\bibfnamefont {V.~D.}\ \bibnamefont
  {Toneev}}, \bibinfo {author} {\bibfnamefont {W.}~\bibnamefont {Cassing}},
  \bibinfo {author} {\bibfnamefont {E.~L.}\ \bibnamefont {Bratkovskaya}},
  \bibinfo {author} {\bibfnamefont {V.~P.}\ \bibnamefont {Konchakovski}}, \
  and\ \bibinfo {author} {\bibfnamefont {S.~A.}\ \bibnamefont {Voloshin}},\
  }\href {\doibase 10.1103/PhysRevC.83.054911} {\bibfield  {journal} {\bibinfo
  {journal} {Phys. Rev. C}\ }\textbf {\bibinfo {volume} {83}},\ \bibinfo
  {pages} {054911} (\bibinfo {year} {2011})},\ \Eprint
  {http://arxiv.org/abs/1103.4239} {arXiv:1103.4239 [nucl-th]} \BibitemShut
  {NoStop}%
\bibitem [{\citenamefont {Bzdak}\ and\ \citenamefont
  {Skokov}(2012)}]{Bzdak:2011yy}%
  \BibitemOpen
  \bibfield  {author} {\bibinfo {author} {\bibfnamefont {A.}~\bibnamefont
  {Bzdak}}\ and\ \bibinfo {author} {\bibfnamefont {V.}~\bibnamefont {Skokov}},\
  }\href {\doibase 10.1016/j.physletb.2012.02.065} {\bibfield  {journal}
  {\bibinfo  {journal} {Phys. Lett. B}\ }\textbf {\bibinfo {volume} {710}},\
  \bibinfo {pages} {171} (\bibinfo {year} {2012})},\ \Eprint
  {http://arxiv.org/abs/1111.1949} {arXiv:1111.1949 [hep-ph]} \BibitemShut
  {NoStop}%
\bibitem [{\citenamefont {Deng}\ and\ \citenamefont
  {Huang}(2012)}]{Deng:2012pc}%
  \BibitemOpen
  \bibfield  {author} {\bibinfo {author} {\bibfnamefont {W.-T.}\ \bibnamefont
  {Deng}}\ and\ \bibinfo {author} {\bibfnamefont {X.-G.}\ \bibnamefont
  {Huang}},\ }\href {\doibase 10.1103/PhysRevC.85.044907} {\bibfield  {journal}
  {\bibinfo  {journal} {Phys. Rev. C}\ }\textbf {\bibinfo {volume} {85}},\
  \bibinfo {pages} {044907} (\bibinfo {year} {2012})},\ \Eprint
  {http://arxiv.org/abs/1201.5108} {arXiv:1201.5108 [nucl-th]} \BibitemShut
  {NoStop}%
\bibitem [{\citenamefont {Mo}\ \emph {et~al.}(2013)\citenamefont {Mo},
  \citenamefont {Feng},\ and\ \citenamefont {Shi}}]{Mo:2013qya}%
  \BibitemOpen
  \bibfield  {author} {\bibinfo {author} {\bibfnamefont {Y.-J.}\ \bibnamefont
  {Mo}}, \bibinfo {author} {\bibfnamefont {S.-Q.}\ \bibnamefont {Feng}}, \ and\
  \bibinfo {author} {\bibfnamefont {Y.-F.}\ \bibnamefont {Shi}},\ }\href
  {\doibase 10.1103/PhysRevC.88.024901} {\bibfield  {journal} {\bibinfo
  {journal} {Phys. Rev. C}\ }\textbf {\bibinfo {volume} {88}},\ \bibinfo
  {pages} {024901} (\bibinfo {year} {2013})},\ \Eprint
  {http://arxiv.org/abs/1308.4289} {arXiv:1308.4289 [hep-ph]} \BibitemShut
  {NoStop}%
\bibitem [{\citenamefont {Zhong}\ \emph {et~al.}(2014)\citenamefont {Zhong},
  \citenamefont {Yang}, \citenamefont {Cai},\ and\ \citenamefont
  {Feng}}]{Zhong:2014cda}%
  \BibitemOpen
  \bibfield  {author} {\bibinfo {author} {\bibfnamefont {Y.}~\bibnamefont
  {Zhong}}, \bibinfo {author} {\bibfnamefont {C.-B.}\ \bibnamefont {Yang}},
  \bibinfo {author} {\bibfnamefont {X.}~\bibnamefont {Cai}}, \ and\ \bibinfo
  {author} {\bibfnamefont {S.-Q.}\ \bibnamefont {Feng}},\ }\href {\doibase
  10.1155/2014/193039} {\bibfield  {journal} {\bibinfo  {journal} {Adv. High
  Energy Phys.}\ }\textbf {\bibinfo {volume} {2014}},\ \bibinfo {pages}
  {193039} (\bibinfo {year} {2014})},\ \Eprint {http://arxiv.org/abs/1408.5694}
  {arXiv:1408.5694 [hep-ph]} \BibitemShut {NoStop}%
\bibitem [{\citenamefont {Feng}\ \emph {et~al.}(2018)\citenamefont {Feng},
  \citenamefont {Pei}, \citenamefont {Sun}, \citenamefont {Zhong},\ and\
  \citenamefont {Yin}}]{Feng:2016srp}%
  \BibitemOpen
  \bibfield  {author} {\bibinfo {author} {\bibfnamefont {S.-Q.}\ \bibnamefont
  {Feng}}, \bibinfo {author} {\bibfnamefont {L.}~\bibnamefont {Pei}}, \bibinfo
  {author} {\bibfnamefont {F.}~\bibnamefont {Sun}}, \bibinfo {author}
  {\bibfnamefont {Y.}~\bibnamefont {Zhong}}, \ and\ \bibinfo {author}
  {\bibfnamefont {Z.-B.}\ \bibnamefont {Yin}},\ }\href {\doibase
  10.1088/1674-1137/42/5/054102} {\bibfield  {journal} {\bibinfo  {journal}
  {Chin. Phys. C}\ }\textbf {\bibinfo {volume} {42}},\ \bibinfo {pages}
  {054102} (\bibinfo {year} {2018})},\ \Eprint
  {http://arxiv.org/abs/1609.07550} {arXiv:1609.07550 [nucl-th]} \BibitemShut
  {NoStop}%
\bibitem [{\citenamefont {Kharzeev}\ \emph {et~al.}(2008)\citenamefont
  {Kharzeev}, \citenamefont {McLerran},\ and\ \citenamefont
  {Warringa}}]{Kharzeev:2007jp}%
  \BibitemOpen
  \bibfield  {author} {\bibinfo {author} {\bibfnamefont {D.~E.}\ \bibnamefont
  {Kharzeev}}, \bibinfo {author} {\bibfnamefont {L.~D.}\ \bibnamefont
  {McLerran}}, \ and\ \bibinfo {author} {\bibfnamefont {H.~J.}\ \bibnamefont
  {Warringa}},\ }\href {\doibase 10.1016/j.nuclphysa.2008.02.298} {\bibfield
  {journal} {\bibinfo  {journal} {Nucl. Phys. A}\ }\textbf {\bibinfo {volume}
  {803}},\ \bibinfo {pages} {227} (\bibinfo {year} {2008})},\ \Eprint
  {http://arxiv.org/abs/0711.0950} {arXiv:0711.0950 [hep-ph]} \BibitemShut
  {NoStop}%
\bibitem [{\citenamefont {Kharzeev}\ \emph {et~al.}(2013)\citenamefont
  {Kharzeev}, \citenamefont {Landsteiner}, \citenamefont {Schmitt},\ and\
  \citenamefont {Yee}}]{Kharzeev:2012ph}%
  \BibitemOpen
  \bibfield  {author} {\bibinfo {author} {\bibfnamefont {D.~E.}\ \bibnamefont
  {Kharzeev}}, \bibinfo {author} {\bibfnamefont {K.}~\bibnamefont
  {Landsteiner}}, \bibinfo {author} {\bibfnamefont {A.}~\bibnamefont
  {Schmitt}}, \ and\ \bibinfo {author} {\bibfnamefont {H.-U.}\ \bibnamefont
  {Yee}},\ }\href {\doibase 10.1007/978-3-642-37305-3_1} {\bibfield  {journal}
  {\bibinfo  {journal} {Lect. Notes Phys.}\ }\textbf {\bibinfo {volume}
  {871}},\ \bibinfo {pages} {1} (\bibinfo {year} {2013})},\ \Eprint
  {http://arxiv.org/abs/1211.6245} {arXiv:1211.6245 [hep-ph]} \BibitemShut
  {NoStop}%
\bibitem [{\citenamefont {Zakharov}(2013)}]{Zakharov:2012vv}%
  \BibitemOpen
  \bibfield  {author} {\bibinfo {author} {\bibfnamefont {V.~I.}\ \bibnamefont
  {Zakharov}},\ }\href {\doibase 10.1007/978-3-642-37305-3_11} {\bibfield
  {journal} {\bibinfo  {journal} {Lect. Notes Phys.}\ }\textbf {\bibinfo
  {volume} {871}},\ \bibinfo {pages} {295} (\bibinfo {year} {2013})},\ \Eprint
  {http://arxiv.org/abs/1210.2186} {arXiv:1210.2186 [hep-ph]} \BibitemShut
  {NoStop}%
\bibitem [{\citenamefont {Kharzeev}(2014)}]{Kharzeev:2013ffa}%
  \BibitemOpen
  \bibfield  {author} {\bibinfo {author} {\bibfnamefont {D.~E.}\ \bibnamefont
  {Kharzeev}},\ }\href {\doibase 10.1016/j.ppnp.2014.01.002} {\bibfield
  {journal} {\bibinfo  {journal} {Prog. Part. Nucl. Phys.}\ }\textbf {\bibinfo
  {volume} {75}},\ \bibinfo {pages} {133} (\bibinfo {year} {2014})},\ \Eprint
  {http://arxiv.org/abs/1312.3348} {arXiv:1312.3348 [hep-ph]} \BibitemShut
  {NoStop}%
\bibitem [{\citenamefont {Liao}(2015)}]{Liao:2014ava}%
  \BibitemOpen
  \bibfield  {author} {\bibinfo {author} {\bibfnamefont {J.}~\bibnamefont
  {Liao}},\ }\href {\doibase 10.1007/s12043-015-0984-x} {\bibfield  {journal}
  {\bibinfo  {journal} {Pramana}\ }\textbf {\bibinfo {volume} {84}},\ \bibinfo
  {pages} {901} (\bibinfo {year} {2015})},\ \Eprint
  {http://arxiv.org/abs/1401.2500} {arXiv:1401.2500 [hep-ph]} \BibitemShut
  {NoStop}%
\bibitem [{\citenamefont {Guo}\ \emph {et~al.}(2019)\citenamefont {Guo},
  \citenamefont {Shi}, \citenamefont {Feng},\ and\ \citenamefont
  {Liao}}]{Guo:2019joy}%
  \BibitemOpen
  \bibfield  {author} {\bibinfo {author} {\bibfnamefont {Y.}~\bibnamefont
  {Guo}}, \bibinfo {author} {\bibfnamefont {S.}~\bibnamefont {Shi}}, \bibinfo
  {author} {\bibfnamefont {S.}~\bibnamefont {Feng}}, \ and\ \bibinfo {author}
  {\bibfnamefont {J.}~\bibnamefont {Liao}},\ }\href {\doibase
  10.1016/j.physletb.2019.134929} {\bibfield  {journal} {\bibinfo  {journal}
  {Phys. Lett. B}\ }\textbf {\bibinfo {volume} {798}},\ \bibinfo {pages}
  {134929} (\bibinfo {year} {2019})},\ \Eprint
  {http://arxiv.org/abs/1905.12613} {arXiv:1905.12613 [nucl-th]} \BibitemShut
  {NoStop}%
\bibitem [{\citenamefont {She}\ \emph {et~al.}(2018)\citenamefont {She},
  \citenamefont {Feng}, \citenamefont {Zhong},\ and\ \citenamefont
  {Yin}}]{She:2017icp}%
  \BibitemOpen
  \bibfield  {author} {\bibinfo {author} {\bibfnamefont {D.}~\bibnamefont
  {She}}, \bibinfo {author} {\bibfnamefont {S.-Q.}\ \bibnamefont {Feng}},
  \bibinfo {author} {\bibfnamefont {Y.}~\bibnamefont {Zhong}}, \ and\ \bibinfo
  {author} {\bibfnamefont {Z.-B.}\ \bibnamefont {Yin}},\ }\href {\doibase
  10.1140/epja/i2018-12481-x} {\bibfield  {journal} {\bibinfo  {journal} {Eur.
  Phys. J. A}\ }\textbf {\bibinfo {volume} {54}},\ \bibinfo {pages} {48}
  (\bibinfo {year} {2018})},\ \Eprint {http://arxiv.org/abs/1709.04662}
  {arXiv:1709.04662 [hep-ph]} \BibitemShut {NoStop}%
\bibitem [{\citenamefont {Andreev}\ and\ \citenamefont
  {Zakharov}(2007{\natexlab{a}})}]{Andreev:2006nw}%
  \BibitemOpen
  \bibfield  {author} {\bibinfo {author} {\bibfnamefont {O.}~\bibnamefont
  {Andreev}}\ and\ \bibinfo {author} {\bibfnamefont {V.~I.}\ \bibnamefont
  {Zakharov}},\ }\href {\doibase 10.1088/1126-6708/2007/04/100} {\bibfield
  {journal} {\bibinfo  {journal} {JHEP}\ }\textbf {\bibinfo {volume} {04}},\
  \bibinfo {pages} {100} (\bibinfo {year} {2007}{\natexlab{a}})},\ \Eprint
  {http://arxiv.org/abs/hep-ph/0611304} {arXiv:hep-ph/0611304} \BibitemShut
  {NoStop}%
\bibitem [{\citenamefont {Chen}\ \emph {et~al.}(2018)\citenamefont {Chen},
  \citenamefont {Feng}, \citenamefont {Shi},\ and\ \citenamefont
  {Zhong}}]{Chen:2017lsf}%
  \BibitemOpen
  \bibfield  {author} {\bibinfo {author} {\bibfnamefont {X.}~\bibnamefont
  {Chen}}, \bibinfo {author} {\bibfnamefont {S.-Q.}\ \bibnamefont {Feng}},
  \bibinfo {author} {\bibfnamefont {Y.-F.}\ \bibnamefont {Shi}}, \ and\
  \bibinfo {author} {\bibfnamefont {Y.}~\bibnamefont {Zhong}},\ }\href
  {\doibase 10.1103/PhysRevD.97.066015} {\bibfield  {journal} {\bibinfo
  {journal} {Phys. Rev. D}\ }\textbf {\bibinfo {volume} {97}},\ \bibinfo
  {pages} {066015} (\bibinfo {year} {2018})},\ \Eprint
  {http://arxiv.org/abs/1710.00465} {arXiv:1710.00465 [hep-ph]} \BibitemShut
  {NoStop}%
\bibitem [{\citenamefont {Bohra}\ \emph {et~al.}(2020)\citenamefont {Bohra},
  \citenamefont {Dudal}, \citenamefont {Hajilou},\ and\ \citenamefont
  {Mahapatra}}]{Bohra:2019ebj}%
  \BibitemOpen
  \bibfield  {author} {\bibinfo {author} {\bibfnamefont {H.}~\bibnamefont
  {Bohra}}, \bibinfo {author} {\bibfnamefont {D.}~\bibnamefont {Dudal}},
  \bibinfo {author} {\bibfnamefont {A.}~\bibnamefont {Hajilou}}, \ and\
  \bibinfo {author} {\bibfnamefont {S.}~\bibnamefont {Mahapatra}},\ }\href
  {\doibase 10.1016/j.physletb.2019.135184} {\bibfield  {journal} {\bibinfo
  {journal} {Phys. Lett. B}\ }\textbf {\bibinfo {volume} {801}},\ \bibinfo
  {pages} {135184} (\bibinfo {year} {2020})},\ \Eprint
  {http://arxiv.org/abs/1907.01852} {arXiv:1907.01852 [hep-th]} \BibitemShut
  {NoStop}%
\bibitem [{\citenamefont {Rodrigues}\ \emph
  {et~al.}(2018{\natexlab{a}})\citenamefont {Rodrigues}, \citenamefont
  {Folco~Capossoli},\ and\ \citenamefont {Boschi-Filho}}]{Rodrigues:2017cha}%
  \BibitemOpen
  \bibfield  {author} {\bibinfo {author} {\bibfnamefont {D.~M.}\ \bibnamefont
  {Rodrigues}}, \bibinfo {author} {\bibfnamefont {E.}~\bibnamefont
  {Folco~Capossoli}}, \ and\ \bibinfo {author} {\bibfnamefont {H.}~\bibnamefont
  {Boschi-Filho}},\ }\href {\doibase 10.1016/j.physletb.2018.02.049} {\bibfield
   {journal} {\bibinfo  {journal} {Phys. Lett. B}\ }\textbf {\bibinfo {volume}
  {780}},\ \bibinfo {pages} {37} (\bibinfo {year} {2018}{\natexlab{a}})},\
  \Eprint {http://arxiv.org/abs/1709.09258} {arXiv:1709.09258 [hep-th]}
  \BibitemShut {NoStop}%
\bibitem [{\citenamefont {Rodrigues}\ \emph
  {et~al.}(2018{\natexlab{b}})\citenamefont {Rodrigues}, \citenamefont
  {Folco~Capossoli},\ and\ \citenamefont {Boschi-Filho}}]{Rodrigues:2017iqi}%
  \BibitemOpen
  \bibfield  {author} {\bibinfo {author} {\bibfnamefont {D.~M.}\ \bibnamefont
  {Rodrigues}}, \bibinfo {author} {\bibfnamefont {E.}~\bibnamefont
  {Folco~Capossoli}}, \ and\ \bibinfo {author} {\bibfnamefont {H.}~\bibnamefont
  {Boschi-Filho}},\ }\href {\doibase 10.1103/PhysRevD.97.126001} {\bibfield
  {journal} {\bibinfo  {journal} {Phys. Rev. D}\ }\textbf {\bibinfo {volume}
  {97}},\ \bibinfo {pages} {126001} (\bibinfo {year} {2018}{\natexlab{b}})},\
  \Eprint {http://arxiv.org/abs/1710.07310} {arXiv:1710.07310 [hep-th]}
  \BibitemShut {NoStop}%
\bibitem [{\citenamefont {Rodrigues}\ \emph
  {et~al.}(2018{\natexlab{c}})\citenamefont {Rodrigues}, \citenamefont {Li},
  \citenamefont {Folco~Capossoli},\ and\ \citenamefont
  {Boschi-Filho}}]{Rodrigues:2018pep}%
  \BibitemOpen
  \bibfield  {author} {\bibinfo {author} {\bibfnamefont {D.~M.}\ \bibnamefont
  {Rodrigues}}, \bibinfo {author} {\bibfnamefont {D.}~\bibnamefont {Li}},
  \bibinfo {author} {\bibfnamefont {E.}~\bibnamefont {Folco~Capossoli}}, \ and\
  \bibinfo {author} {\bibfnamefont {H.}~\bibnamefont {Boschi-Filho}},\ }\href
  {\doibase 10.1103/PhysRevD.98.106007} {\bibfield  {journal} {\bibinfo
  {journal} {Phys. Rev. D}\ }\textbf {\bibinfo {volume} {98}},\ \bibinfo
  {pages} {106007} (\bibinfo {year} {2018}{\natexlab{c}})},\ \Eprint
  {http://arxiv.org/abs/1807.11822} {arXiv:1807.11822 [hep-th]} \BibitemShut
  {NoStop}%
\bibitem [{\citenamefont {McInnes}(2016)}]{McInnes:2015kec}%
  \BibitemOpen
  \bibfield  {author} {\bibinfo {author} {\bibfnamefont {B.}~\bibnamefont
  {McInnes}},\ }\href {\doibase 10.1016/j.nuclphysb.2016.02.027} {\bibfield
  {journal} {\bibinfo  {journal} {Nucl. Phys. B}\ }\textbf {\bibinfo {volume}
  {906}},\ \bibinfo {pages} {40} (\bibinfo {year} {2016})},\ \Eprint
  {http://arxiv.org/abs/1511.05293} {arXiv:1511.05293 [hep-th]} \BibitemShut
  {NoStop}%
\bibitem [{\citenamefont {Gursoy}\ \emph {et~al.}(2018)\citenamefont {Gursoy},
  \citenamefont {Jarvinen},\ and\ \citenamefont {Nijs}}]{Gursoy:2017wzz}%
  \BibitemOpen
  \bibfield  {author} {\bibinfo {author} {\bibfnamefont {U.}~\bibnamefont
  {Gursoy}}, \bibinfo {author} {\bibfnamefont {M.}~\bibnamefont {Jarvinen}}, \
  and\ \bibinfo {author} {\bibfnamefont {G.}~\bibnamefont {Nijs}},\ }\href
  {\doibase 10.1103/PhysRevLett.120.242002} {\bibfield  {journal} {\bibinfo
  {journal} {Phys. Rev. Lett.}\ }\textbf {\bibinfo {volume} {120}},\ \bibinfo
  {pages} {242002} (\bibinfo {year} {2018})},\ \Eprint
  {http://arxiv.org/abs/1707.00872} {arXiv:1707.00872 [hep-th]} \BibitemShut
  {NoStop}%
\bibitem [{\citenamefont {G\"ursoy}\ \emph {et~al.}(2017)\citenamefont
  {G\"ursoy}, \citenamefont {Iatrakis}, \citenamefont {J\"arvinen},\ and\
  \citenamefont {Nijs}}]{Gursoy:2016ofp}%
  \BibitemOpen
  \bibfield  {author} {\bibinfo {author} {\bibfnamefont {U.}~\bibnamefont
  {G\"ursoy}}, \bibinfo {author} {\bibfnamefont {I.}~\bibnamefont {Iatrakis}},
  \bibinfo {author} {\bibfnamefont {M.}~\bibnamefont {J\"arvinen}}, \ and\
  \bibinfo {author} {\bibfnamefont {G.}~\bibnamefont {Nijs}},\ }\href {\doibase
  10.1007/JHEP03(2017)053} {\bibfield  {journal} {\bibinfo  {journal} {JHEP}\
  }\textbf {\bibinfo {volume} {03}},\ \bibinfo {pages} {053} (\bibinfo {year}
  {2017})},\ \Eprint {http://arxiv.org/abs/1611.06339} {arXiv:1611.06339
  [hep-th]} \BibitemShut {NoStop}%
\bibitem [{\citenamefont {Dudal}\ and\ \citenamefont
  {Mertens}(2018)}]{Dudal:2018rki}%
  \BibitemOpen
  \bibfield  {author} {\bibinfo {author} {\bibfnamefont {D.}~\bibnamefont
  {Dudal}}\ and\ \bibinfo {author} {\bibfnamefont {T.~G.}\ \bibnamefont
  {Mertens}},\ }\href {\doibase 10.1103/PhysRevD.97.054035} {\bibfield
  {journal} {\bibinfo  {journal} {Phys. Rev. D}\ }\textbf {\bibinfo {volume}
  {97}},\ \bibinfo {pages} {054035} (\bibinfo {year} {2018})},\ \Eprint
  {http://arxiv.org/abs/1802.02805} {arXiv:1802.02805 [hep-th]} \BibitemShut
  {NoStop}%
\bibitem [{\citenamefont {Bhattacharyya}\ \emph
  {et~al.}(2008{\natexlab{a}})\citenamefont {Bhattacharyya}, \citenamefont
  {Hubeny}, \citenamefont {Minwalla},\ and\ \citenamefont
  {Rangamani}}]{Bhattacharyya:2007vjd}%
  \BibitemOpen
  \bibfield  {author} {\bibinfo {author} {\bibfnamefont {S.}~\bibnamefont
  {Bhattacharyya}}, \bibinfo {author} {\bibfnamefont {V.~E.}\ \bibnamefont
  {Hubeny}}, \bibinfo {author} {\bibfnamefont {S.}~\bibnamefont {Minwalla}}, \
  and\ \bibinfo {author} {\bibfnamefont {M.}~\bibnamefont {Rangamani}},\ }\href
  {\doibase 10.1088/1126-6708/2008/02/045} {\bibfield  {journal} {\bibinfo
  {journal} {JHEP}\ }\textbf {\bibinfo {volume} {02}},\ \bibinfo {pages} {045}
  (\bibinfo {year} {2008}{\natexlab{a}})},\ \Eprint
  {http://arxiv.org/abs/0712.2456} {arXiv:0712.2456 [hep-th]} \BibitemShut
  {NoStop}%
\bibitem [{\citenamefont {Bhattacharyya}\ \emph
  {et~al.}(2008{\natexlab{b}})\citenamefont {Bhattacharyya}, \citenamefont
  {Hubeny}, \citenamefont {Loganayagam}, \citenamefont {Mandal}, \citenamefont
  {Minwalla}, \citenamefont {Morita}, \citenamefont {Rangamani},\ and\
  \citenamefont {Reall}}]{Bhattacharyya:2008xc}%
  \BibitemOpen
  \bibfield  {author} {\bibinfo {author} {\bibfnamefont {S.}~\bibnamefont
  {Bhattacharyya}}, \bibinfo {author} {\bibfnamefont {V.~E.}\ \bibnamefont
  {Hubeny}}, \bibinfo {author} {\bibfnamefont {R.}~\bibnamefont {Loganayagam}},
  \bibinfo {author} {\bibfnamefont {G.}~\bibnamefont {Mandal}}, \bibinfo
  {author} {\bibfnamefont {S.}~\bibnamefont {Minwalla}}, \bibinfo {author}
  {\bibfnamefont {T.}~\bibnamefont {Morita}}, \bibinfo {author} {\bibfnamefont
  {M.}~\bibnamefont {Rangamani}}, \ and\ \bibinfo {author} {\bibfnamefont
  {H.~S.}\ \bibnamefont {Reall}},\ }\href {\doibase
  10.1088/1126-6708/2008/06/055} {\bibfield  {journal} {\bibinfo  {journal}
  {JHEP}\ }\textbf {\bibinfo {volume} {06}},\ \bibinfo {pages} {055} (\bibinfo
  {year} {2008}{\natexlab{b}})},\ \Eprint {http://arxiv.org/abs/0803.2526}
  {arXiv:0803.2526 [hep-th]} \BibitemShut {NoStop}%
\bibitem [{\citenamefont {Loganayagam}(2008)}]{Loganayagam:2008is}%
  \BibitemOpen
  \bibfield  {author} {\bibinfo {author} {\bibfnamefont {R.}~\bibnamefont
  {Loganayagam}},\ }\href {\doibase 10.1088/1126-6708/2008/05/087} {\bibfield
  {journal} {\bibinfo  {journal} {JHEP}\ }\textbf {\bibinfo {volume} {05}},\
  \bibinfo {pages} {087} (\bibinfo {year} {2008})},\ \Eprint
  {http://arxiv.org/abs/0801.3701} {arXiv:0801.3701 [hep-th]} \BibitemShut
  {NoStop}%
\bibitem [{\citenamefont {Chapman}\ \emph {et~al.}(2012)\citenamefont
  {Chapman}, \citenamefont {Neiman},\ and\ \citenamefont
  {Oz}}]{Chapman:2012my}%
  \BibitemOpen
  \bibfield  {author} {\bibinfo {author} {\bibfnamefont {S.}~\bibnamefont
  {Chapman}}, \bibinfo {author} {\bibfnamefont {Y.}~\bibnamefont {Neiman}}, \
  and\ \bibinfo {author} {\bibfnamefont {Y.}~\bibnamefont {Oz}},\ }\href
  {\doibase 10.1007/JHEP07(2012)128} {\bibfield  {journal} {\bibinfo  {journal}
  {JHEP}\ }\textbf {\bibinfo {volume} {07}},\ \bibinfo {pages} {128} (\bibinfo
  {year} {2012})},\ \Eprint {http://arxiv.org/abs/1202.2469} {arXiv:1202.2469
  [hep-th]} \BibitemShut {NoStop}%
\bibitem [{\citenamefont {Erdmenger}\ \emph {et~al.}(2009)\citenamefont
  {Erdmenger}, \citenamefont {Haack}, \citenamefont {Kaminski},\ and\
  \citenamefont {Yarom}}]{Erdmenger:2008rm}%
  \BibitemOpen
  \bibfield  {author} {\bibinfo {author} {\bibfnamefont {J.}~\bibnamefont
  {Erdmenger}}, \bibinfo {author} {\bibfnamefont {M.}~\bibnamefont {Haack}},
  \bibinfo {author} {\bibfnamefont {M.}~\bibnamefont {Kaminski}}, \ and\
  \bibinfo {author} {\bibfnamefont {A.}~\bibnamefont {Yarom}},\ }\href
  {\doibase 10.1088/1126-6708/2009/01/055} {\bibfield  {journal} {\bibinfo
  {journal} {JHEP}\ }\textbf {\bibinfo {volume} {01}},\ \bibinfo {pages} {055}
  (\bibinfo {year} {2009})},\ \Eprint {http://arxiv.org/abs/0809.2488}
  {arXiv:0809.2488 [hep-th]} \BibitemShut {NoStop}%
\bibitem [{\citenamefont {Banerjee}\ \emph {et~al.}(2011)\citenamefont
  {Banerjee}, \citenamefont {Bhattacharya}, \citenamefont {Bhattacharyya},
  \citenamefont {Dutta}, \citenamefont {Loganayagam},\ and\ \citenamefont
  {Surowka}}]{Banerjee:2008th}%
  \BibitemOpen
  \bibfield  {author} {\bibinfo {author} {\bibfnamefont {N.}~\bibnamefont
  {Banerjee}}, \bibinfo {author} {\bibfnamefont {J.}~\bibnamefont
  {Bhattacharya}}, \bibinfo {author} {\bibfnamefont {S.}~\bibnamefont
  {Bhattacharyya}}, \bibinfo {author} {\bibfnamefont {S.}~\bibnamefont
  {Dutta}}, \bibinfo {author} {\bibfnamefont {R.}~\bibnamefont {Loganayagam}},
  \ and\ \bibinfo {author} {\bibfnamefont {P.}~\bibnamefont {Surowka}},\ }\href
  {\doibase 10.1007/JHEP01(2011)094} {\bibfield  {journal} {\bibinfo  {journal}
  {JHEP}\ }\textbf {\bibinfo {volume} {01}},\ \bibinfo {pages} {094} (\bibinfo
  {year} {2011})},\ \Eprint {http://arxiv.org/abs/0809.2596} {arXiv:0809.2596
  [hep-th]} \BibitemShut {NoStop}%
\bibitem [{\citenamefont {Megias}\ and\ \citenamefont
  {Pena-Benitez}(2013)}]{Megias:2013joa}%
  \BibitemOpen
  \bibfield  {author} {\bibinfo {author} {\bibfnamefont {E.}~\bibnamefont
  {Megias}}\ and\ \bibinfo {author} {\bibfnamefont {F.}~\bibnamefont
  {Pena-Benitez}},\ }\href {\doibase 10.1007/JHEP05(2013)115} {\bibfield
  {journal} {\bibinfo  {journal} {JHEP}\ }\textbf {\bibinfo {volume} {05}},\
  \bibinfo {pages} {115} (\bibinfo {year} {2013})},\ \Eprint
  {http://arxiv.org/abs/1304.5529} {arXiv:1304.5529 [hep-th]} \BibitemShut
  {NoStop}%
\bibitem [{\citenamefont {Son}\ and\ \citenamefont
  {Surowka}(2009)}]{Son:2009tf}%
  \BibitemOpen
  \bibfield  {author} {\bibinfo {author} {\bibfnamefont {D.~T.}\ \bibnamefont
  {Son}}\ and\ \bibinfo {author} {\bibfnamefont {P.}~\bibnamefont {Surowka}},\
  }\href {\doibase 10.1103/PhysRevLett.103.191601} {\bibfield  {journal}
  {\bibinfo  {journal} {Phys. Rev. Lett.}\ }\textbf {\bibinfo {volume} {103}},\
  \bibinfo {pages} {191601} (\bibinfo {year} {2009})},\ \Eprint
  {http://arxiv.org/abs/0906.5044} {arXiv:0906.5044 [hep-th]} \BibitemShut
  {NoStop}%
\bibitem [{\citenamefont {Bhattacharyya}\ \emph {et~al.}(2009)\citenamefont
  {Bhattacharyya}, \citenamefont {Loganayagam}, \citenamefont {Minwalla},
  \citenamefont {Nampuri}, \citenamefont {Trivedi},\ and\ \citenamefont
  {Wadia}}]{Bhattacharyya:2008ji}%
  \BibitemOpen
  \bibfield  {author} {\bibinfo {author} {\bibfnamefont {S.}~\bibnamefont
  {Bhattacharyya}}, \bibinfo {author} {\bibfnamefont {R.}~\bibnamefont
  {Loganayagam}}, \bibinfo {author} {\bibfnamefont {S.}~\bibnamefont
  {Minwalla}}, \bibinfo {author} {\bibfnamefont {S.}~\bibnamefont {Nampuri}},
  \bibinfo {author} {\bibfnamefont {S.~P.}\ \bibnamefont {Trivedi}}, \ and\
  \bibinfo {author} {\bibfnamefont {S.~R.}\ \bibnamefont {Wadia}},\ }\href
  {\doibase 10.1088/1126-6708/2009/02/018} {\bibfield  {journal} {\bibinfo
  {journal} {JHEP}\ }\textbf {\bibinfo {volume} {02}},\ \bibinfo {pages} {018}
  (\bibinfo {year} {2009})},\ \Eprint {http://arxiv.org/abs/0806.0006}
  {arXiv:0806.0006 [hep-th]} \BibitemShut {NoStop}%
\bibitem [{\citenamefont {Andreev}(2006)}]{Andreev:2006vy}%
  \BibitemOpen
  \bibfield  {author} {\bibinfo {author} {\bibfnamefont {O.}~\bibnamefont
  {Andreev}},\ }\href {\doibase 10.1103/PhysRevD.73.107901} {\bibfield
  {journal} {\bibinfo  {journal} {Phys. Rev. D}\ }\textbf {\bibinfo {volume}
  {73}},\ \bibinfo {pages} {107901} (\bibinfo {year} {2006})},\ \Eprint
  {http://arxiv.org/abs/hep-th/0603170} {arXiv:hep-th/0603170} \BibitemShut
  {NoStop}%
\bibitem [{\citenamefont {Andreev}\ and\ \citenamefont
  {Zakharov}(2006)}]{Andreev:2006ct}%
  \BibitemOpen
  \bibfield  {author} {\bibinfo {author} {\bibfnamefont {O.}~\bibnamefont
  {Andreev}}\ and\ \bibinfo {author} {\bibfnamefont {V.~I.}\ \bibnamefont
  {Zakharov}},\ }\href {\doibase 10.1103/PhysRevD.74.025023} {\bibfield
  {journal} {\bibinfo  {journal} {Phys. Rev. D}\ }\textbf {\bibinfo {volume}
  {74}},\ \bibinfo {pages} {025023} (\bibinfo {year} {2006})},\ \Eprint
  {http://arxiv.org/abs/hep-ph/0604204} {arXiv:hep-ph/0604204} \BibitemShut
  {NoStop}%
\bibitem [{\citenamefont {Andreev}(2010)}]{Andreev:2010bv}%
  \BibitemOpen
  \bibfield  {author} {\bibinfo {author} {\bibfnamefont {O.}~\bibnamefont
  {Andreev}},\ }\href {\doibase 10.1103/PhysRevD.81.087901} {\bibfield
  {journal} {\bibinfo  {journal} {Phys. Rev. D}\ }\textbf {\bibinfo {volume}
  {81}},\ \bibinfo {pages} {087901} (\bibinfo {year} {2010})},\ \Eprint
  {http://arxiv.org/abs/1001.4414} {arXiv:1001.4414 [hep-ph]} \BibitemShut
  {NoStop}%
\bibitem [{\citenamefont {Andreev}\ and\ \citenamefont
  {Zakharov}(2007{\natexlab{b}})}]{Andreev:2006eh}%
  \BibitemOpen
  \bibfield  {author} {\bibinfo {author} {\bibfnamefont {O.}~\bibnamefont
  {Andreev}}\ and\ \bibinfo {author} {\bibfnamefont {V.~I.}\ \bibnamefont
  {Zakharov}},\ }\href {\doibase 10.1016/j.physletb.2007.01.002} {\bibfield
  {journal} {\bibinfo  {journal} {Phys. Lett. B}\ }\textbf {\bibinfo {volume}
  {645}},\ \bibinfo {pages} {437} (\bibinfo {year} {2007}{\natexlab{b}})},\
  \Eprint {http://arxiv.org/abs/hep-ph/0607026} {arXiv:hep-ph/0607026}
  \BibitemShut {NoStop}%
\bibitem [{\citenamefont {Bali}\ \emph {et~al.}(2012)\citenamefont {Bali},
  \citenamefont {Bruckmann}, \citenamefont {Endrodi}, \citenamefont {Fodor},
  \citenamefont {Katz},\ and\ \citenamefont {Schafer}}]{Bali:2012zg}%
  \BibitemOpen
  \bibfield  {author} {\bibinfo {author} {\bibfnamefont {G.~S.}\ \bibnamefont
  {Bali}}, \bibinfo {author} {\bibfnamefont {F.}~\bibnamefont {Bruckmann}},
  \bibinfo {author} {\bibfnamefont {G.}~\bibnamefont {Endrodi}}, \bibinfo
  {author} {\bibfnamefont {Z.}~\bibnamefont {Fodor}}, \bibinfo {author}
  {\bibfnamefont {S.~D.}\ \bibnamefont {Katz}}, \ and\ \bibinfo {author}
  {\bibfnamefont {A.}~\bibnamefont {Schafer}},\ }\href {\doibase
  10.1103/PhysRevD.86.071502} {\bibfield  {journal} {\bibinfo  {journal} {Phys.
  Rev. D}\ }\textbf {\bibinfo {volume} {86}},\ \bibinfo {pages} {071502}
  (\bibinfo {year} {2012})},\ \Eprint {http://arxiv.org/abs/1206.4205}
  {arXiv:1206.4205 [hep-lat]} \BibitemShut {NoStop}%
\bibitem [{\citenamefont {Bali}\ \emph {et~al.}(2014)\citenamefont {Bali},
  \citenamefont {Bruckmann}, \citenamefont {Endr\"odi}, \citenamefont {Katz},\
  and\ \citenamefont {Sch\"afer}}]{Bali:2014kia}%
  \BibitemOpen
  \bibfield  {author} {\bibinfo {author} {\bibfnamefont {G.~S.}\ \bibnamefont
  {Bali}}, \bibinfo {author} {\bibfnamefont {F.}~\bibnamefont {Bruckmann}},
  \bibinfo {author} {\bibfnamefont {G.}~\bibnamefont {Endr\"odi}}, \bibinfo
  {author} {\bibfnamefont {S.~D.}\ \bibnamefont {Katz}}, \ and\ \bibinfo
  {author} {\bibfnamefont {A.}~\bibnamefont {Sch\"afer}},\ }\href {\doibase
  10.1007/JHEP08(2014)177} {\bibfield  {journal} {\bibinfo  {journal} {JHEP}\
  }\textbf {\bibinfo {volume} {08}},\ \bibinfo {pages} {177} (\bibinfo {year}
  {2014})},\ \Eprint {http://arxiv.org/abs/1406.0269} {arXiv:1406.0269
  [hep-lat]} \BibitemShut {NoStop}%
\bibitem [{\citenamefont {Farias}\ \emph {et~al.}(2014)\citenamefont {Farias},
  \citenamefont {Gomes}, \citenamefont {Krein},\ and\ \citenamefont
  {Pinto}}]{Farias:2014eca}%
  \BibitemOpen
  \bibfield  {author} {\bibinfo {author} {\bibfnamefont {R.~L.~S.}\
  \bibnamefont {Farias}}, \bibinfo {author} {\bibfnamefont {K.~P.}\
  \bibnamefont {Gomes}}, \bibinfo {author} {\bibfnamefont {G.~I.}\ \bibnamefont
  {Krein}}, \ and\ \bibinfo {author} {\bibfnamefont {M.~B.}\ \bibnamefont
  {Pinto}},\ }\href {\doibase 10.1103/PhysRevC.90.025203} {\bibfield  {journal}
  {\bibinfo  {journal} {Phys. Rev. C}\ }\textbf {\bibinfo {volume} {90}},\
  \bibinfo {pages} {025203} (\bibinfo {year} {2014})},\ \Eprint
  {http://arxiv.org/abs/1404.3931} {arXiv:1404.3931 [hep-ph]} \BibitemShut
  {NoStop}%
\bibitem [{\citenamefont {Ferreira}\ \emph {et~al.}(2014)\citenamefont
  {Ferreira}, \citenamefont {Costa}, \citenamefont {Louren\c{c}o},
  \citenamefont {Frederico},\ and\ \citenamefont
  {Provid\^encia}}]{Ferreira:2014kpa}%
  \BibitemOpen
  \bibfield  {author} {\bibinfo {author} {\bibfnamefont {M.}~\bibnamefont
  {Ferreira}}, \bibinfo {author} {\bibfnamefont {P.}~\bibnamefont {Costa}},
  \bibinfo {author} {\bibfnamefont {O.}~\bibnamefont {Louren\c{c}o}}, \bibinfo
  {author} {\bibfnamefont {T.}~\bibnamefont {Frederico}}, \ and\ \bibinfo
  {author} {\bibfnamefont {C.}~\bibnamefont {Provid\^encia}},\ }\href {\doibase
  10.1103/PhysRevD.89.116011} {\bibfield  {journal} {\bibinfo  {journal} {Phys.
  Rev. D}\ }\textbf {\bibinfo {volume} {89}},\ \bibinfo {pages} {116011}
  (\bibinfo {year} {2014})},\ \Eprint {http://arxiv.org/abs/1404.5577}
  {arXiv:1404.5577 [hep-ph]} \BibitemShut {NoStop}%
\bibitem [{\citenamefont {Ayala}\ \emph {et~al.}(2014)\citenamefont {Ayala},
  \citenamefont {Loewe}, \citenamefont {Mizher},\ and\ \citenamefont
  {Zamora}}]{Ayala:2014iba}%
  \BibitemOpen
  \bibfield  {author} {\bibinfo {author} {\bibfnamefont {A.}~\bibnamefont
  {Ayala}}, \bibinfo {author} {\bibfnamefont {M.}~\bibnamefont {Loewe}},
  \bibinfo {author} {\bibfnamefont {A.~J.}\ \bibnamefont {Mizher}}, \ and\
  \bibinfo {author} {\bibfnamefont {R.}~\bibnamefont {Zamora}},\ }\href
  {\doibase 10.1103/PhysRevD.90.036001} {\bibfield  {journal} {\bibinfo
  {journal} {Phys. Rev. D}\ }\textbf {\bibinfo {volume} {90}},\ \bibinfo
  {pages} {036001} (\bibinfo {year} {2014})},\ \Eprint
  {http://arxiv.org/abs/1406.3885} {arXiv:1406.3885 [hep-ph]} \BibitemShut
  {NoStop}%
\bibitem [{\citenamefont {Mueller}\ and\ \citenamefont
  {Pawlowski}(2015)}]{Mueller:2015fka}%
  \BibitemOpen
  \bibfield  {author} {\bibinfo {author} {\bibfnamefont {N.}~\bibnamefont
  {Mueller}}\ and\ \bibinfo {author} {\bibfnamefont {J.~M.}\ \bibnamefont
  {Pawlowski}},\ }\href {\doibase 10.1103/PhysRevD.91.116010} {\bibfield
  {journal} {\bibinfo  {journal} {Phys. Rev. D}\ }\textbf {\bibinfo {volume}
  {91}},\ \bibinfo {pages} {116010} (\bibinfo {year} {2015})},\ \Eprint
  {http://arxiv.org/abs/1502.08011} {arXiv:1502.08011 [hep-ph]} \BibitemShut
  {NoStop}%
\bibitem [{\citenamefont {Hashimoto}\ and\ \citenamefont
  {Kharzeev}(2014)}]{Hashimoto:2014fha}%
  \BibitemOpen
  \bibfield  {author} {\bibinfo {author} {\bibfnamefont {K.}~\bibnamefont
  {Hashimoto}}\ and\ \bibinfo {author} {\bibfnamefont {D.~E.}\ \bibnamefont
  {Kharzeev}},\ }\href {\doibase 10.1103/PhysRevD.90.125012} {\bibfield
  {journal} {\bibinfo  {journal} {Phys. Rev. D}\ }\textbf {\bibinfo {volume}
  {90}},\ \bibinfo {pages} {125012} (\bibinfo {year} {2014})},\ \Eprint
  {http://arxiv.org/abs/1411.0618} {arXiv:1411.0618 [hep-th]} \BibitemShut
  {NoStop}%
\bibitem [{\citenamefont {Kaczmarek}\ \emph {et~al.}(2002)\citenamefont
  {Kaczmarek}, \citenamefont {Karsch}, \citenamefont {Petreczky},\ and\
  \citenamefont {Zantow}}]{Kaczmarek:2002mc}%
  \BibitemOpen
  \bibfield  {author} {\bibinfo {author} {\bibfnamefont {O.}~\bibnamefont
  {Kaczmarek}}, \bibinfo {author} {\bibfnamefont {F.}~\bibnamefont {Karsch}},
  \bibinfo {author} {\bibfnamefont {P.}~\bibnamefont {Petreczky}}, \ and\
  \bibinfo {author} {\bibfnamefont {F.}~\bibnamefont {Zantow}},\ }\href
  {\doibase 10.1016/S0370-2693(02)02415-2} {\bibfield  {journal} {\bibinfo
  {journal} {Phys. Lett. B}\ }\textbf {\bibinfo {volume} {543}},\ \bibinfo
  {pages} {41} (\bibinfo {year} {2002})},\ \Eprint
  {http://arxiv.org/abs/hep-lat/0207002} {arXiv:hep-lat/0207002} \BibitemShut
  {NoStop}%
\bibitem [{\citenamefont {Petreczky}\ and\ \citenamefont
  {Petrov}(2004)}]{Petreczky:2004pz}%
  \BibitemOpen
  \bibfield  {author} {\bibinfo {author} {\bibfnamefont {P.}~\bibnamefont
  {Petreczky}}\ and\ \bibinfo {author} {\bibfnamefont {K.}~\bibnamefont
  {Petrov}},\ }\href {\doibase 10.1103/PhysRevD.70.054503} {\bibfield
  {journal} {\bibinfo  {journal} {Phys. Rev. D}\ }\textbf {\bibinfo {volume}
  {70}},\ \bibinfo {pages} {054503} (\bibinfo {year} {2004})},\ \Eprint
  {http://arxiv.org/abs/hep-lat/0405009} {arXiv:hep-lat/0405009} \BibitemShut
  {NoStop}%
\bibitem [{\citenamefont {Kaczmarek}\ and\ \citenamefont
  {Zantow}(2006)}]{Kaczmarek:2005zp}%
  \BibitemOpen
  \bibfield  {author} {\bibinfo {author} {\bibfnamefont {O.}~\bibnamefont
  {Kaczmarek}}\ and\ \bibinfo {author} {\bibfnamefont {F.}~\bibnamefont
  {Zantow}},\ }\href {\doibase 10.22323/1.020.0192} {\bibfield  {journal}
  {\bibinfo  {journal} {PoS}\ }\textbf {\bibinfo {volume} {LAT2005}},\ \bibinfo
  {pages} {192} (\bibinfo {year} {2006})},\ \Eprint
  {http://arxiv.org/abs/hep-lat/0510094} {arXiv:hep-lat/0510094} \BibitemShut
  {NoStop}%
\end{thebibliography}%
\end{document}